\documentclass[%
 aip,
 amsmath,amssymb,
 reprint,%
]{revtex4-1}

\usepackage{graphicx}
\usepackage{dcolumn}
\usepackage{bm}

\usepackage[utf8]{inputenc}
\usepackage[T1]{fontenc}
\usepackage{mathptmx}
\usepackage{etoolbox}
\usepackage{xcolor,soul}
\sethlcolor{yellow}

\makeatletter
\def\@email#1#2{%
 \endgroup
 \patchcmd{\titleblock@produce}
  {\frontmatter@RRAPformat}
  {\frontmatter@RRAPformat{\produce@RRAP{*#1\href{mailto:#2}{#2}}}\frontmatter@RRAPformat}
  {}{}
}%
\makeatother
\begin{document}
\title[A Fuzzy Classification Framework to Identify Equivalent Atoms in Complex Materials and Molecules]{A Fuzzy Classification Framework to\\ Identify Equivalent Atoms in Complex Materials and Molecules}
\author{King Chun Lai}\email{lai@fhi-berlin.mpg.de}
\author{Sebastian Matera}
\author{Christoph Scheurer}
\author{Karsten Reuter}
\affiliation{Fritz-Haber-Institut der Max-Planck-Gesellschaft, Faradayweg 4-6, 14195 Berlin, Germany}

\date{31 May 2023}

\begin{abstract}
The nature of an atom in a bonded structure -- such as in molecules, in nanoparticles or solids, at surfaces or interfaces -- depends on its local atomic environment. In atomic-scale modeling and simulation, identifying groups of atoms with equivalent environments is a frequent task, to gain an understanding of the material function, to interpret experimental results or to simply restrict demanding first-principles calculations. While routine, this task can often be challenging for complex molecules or non-ideal materials with breaks of symmetries or long-range order. To automatize this task, we here present a general machine-learning framework to identify groups of (nearly) equivalent atoms. The initial classification rests on the representation of the local atomic environment through a high-dimensional smooth overlap of atomic positions (SOAP) vector. Recognizing that not least thermal vibrations may lead to deviations from ideal positions, we then achieve a fuzzy classification by mean-shift clustering within a low-dimensional embedded representation of the SOAP points as obtained through multidimensional scaling. The performance of this classification framework is demonstrated for simple aromatic molecules and crystalline Pd surface examples.
\end{abstract}

\maketitle
\section{Introduction}\label{sec:intro}

When bound into molecules or materials, even atoms of the same chemical species can still possess very different properties and functions, e.g. different roles in a chemical reaction. A decisive factor for this is the local atomic environment of the atom, i.e. the relative positions of all other atoms in its vicinity. A natural question is then which of the atoms in one or several different bonded structures are equivalent in terms of this local environment and would correspondingly be attributed similar properties and functions. Indeed, such a grouping of equivalent atoms is common in material science and chemistry. In atomic-scale modeling and simulation it is e.g. central to allocate computational effort to representative atoms of each equivalence class, to structure the data analysis, to select building blocks in material design -- to name but a few of the frequent use cases. A specific application that served as original motivation for this work would, for instance, be adaptive kinetic Monte Carlo (kMC) simulations{\cite{andersen2019kMC,reuter2016ab_initio}}, where transition states of elementary processes need to be computed for every atom in a structure in a potentially huge number of sequential kMC steps. Good starting guesses for the transition states based on recognizing that an atom has a similar local environment to previously calculated cases is there a pivotal efficiency driver.

Now, it is intuitively clear that a small perturbation of the local environment will generally not dramatically change the nature of an atom. Likewise, the nearsightedness of chemical interactions also tells that neighboring atoms further and further away will typically play an ever decreasing role. In practice, the classification of equivalent atoms should therefore be fuzzy, up to such small perturbations and prioritizing close by neighbors. In fact, the resolution, i.e. with up to which differences in their local environment atoms are still classified as equivalent, is a continuous function and the optimum resolution will depend on the bonded structure and the task at hand. For instance, for organic molecules the direct coordination of an atom may already be enough to obtain a qualitative understanding of its function. A carbon atom in an aromatic ring will have very different properties to a carbon atom in an alkyl chain, and a coarse representation of the local environment accounting only for the directly coordinated neighbors, their distances and bond angles would suffice to distinguish the two cases. Similarly, at crystalline metal surfaces, a first distinction is generally made in terms of differently coordinated terrace atoms, step atoms, kink atoms or adatoms. However, depending on the application it may also be necessary to further branch these into sub-classes resolving e.g. the surface orientation (facet), the step type, combinations of multiple chemical species, nearby defects or other increasingly more subtle variations in the local atomic environments.

Traditionally, the grouping into equivalent atoms is performed manually by the researcher and is often merely based on visual inspection of the atomic structure. This approach is obviously laborious and error-prone, and conflicts with increasing interest in high-throughput workflows \cite{yang2022natrevmat,WWL-GPR,HighThroughput_05,HighThroughput_01}, e.g. for catalysis \cite{toyao2019catalysis} or battery interfaces \cite{bhowmik2019battery,steinmann2021battery}, with interest in the generation of large and growing structural databases \cite{stocker2020database,schober2016database}, in global structure optimization problems \cite{Hammer_1,Hammer_2,Hammer_3, MinimaHopping2004, RandomSearch2011}, or in the treatment of complex atomic arrangements, such as nanostructures \cite{banerjee2021nanostructures} or amorphous materials \cite{stegmaier2021AdvMat}. In these tasks, identifying a complete set of equivalent local environments merely by visual inspection would either become a severe limitation or be completely intractable.

To address this issue, we here develop a general machine-learning (ML) framework to automatically identify the groups of (near-)equivalent atoms within any single or any set of bonded structures. These bonded structures may thereby comprise molecules, extended (crystalline or amorphous) materials, as well as their surfaces or interfaces. Emphasis is made to have a simple and continuous control of the resolution in the fuzzy classification. The starting point is to utilize one of the local descriptors \cite{SOAP_Bartok_2013,MBTR_Huo_2017,ACE,descriptor2018,representation_chemrev}, which have been developed during the last years to map the local environment of an atom onto a point in a high-dimensional space $\mathbb{R}^S$. After determining this vector for all atoms in the considered structure(s), we employ clustering on the resulting set of data points to obtain different classes of equivalent environments. Fuzziness is introduced in this approach by specifically employing the double smooth overlap of atomic positions (SOAP) descriptor \cite{SOAP_Bartok_2013}, which naturally emphasizes nearsightedness, and by employing multidimensional scaling (MDS) \cite{MDS_Carroll_1998} to embed the SOAP-points in a lower-dimensional space $\mathbb{R}^{S'}$. This lower-dimensional space is then beneficial to obtain the fuzzy classes of approximately equivalent atoms by mean shift clustering (MSC) \cite{MeanShift_Fukunaga_1975}. Besides the parameters of the SOAP representation, the framework thus has two key hyperparameters to control the resolution of the classification, i.e. the dimensionality of the MDS space and the bandwidth of the MSC. 

We would like to emphasize that algorithms to categorize atoms on the basis of their local environments have been proposed for multiple application purposes before. A few prominent examples are the work of the Hammer group in the context of global geometry optimization \cite{Hammer_1,Hammer_3}, or the work of the Ceriotti group in the context of probabilistic analysis of molecular motifs (PAMM){\cite{ceriotti_PAMM_hydrogen,ceriotti_PAMM2}}. While following analog conceptual steps as e.g. the Hammer workflow, our framework differs in its attempt to avoid any predefinition and system specificity. Rather than {\em a priori} specifying the number of different equivalence classes, this number and the corresponding classes result automatically as a consequence of the chosen resolution as controlled by the MSC bandwidth. With a larger bandwidth, fewer classes will be distinguished and atoms with wider variations in their local environment will still be classified as equivalent. Similarly, rather than imposing system-specific features of the local environment as central for the classification, these features again emerge naturally in the MDS dimensionality reduction step. With lower MDS dimensionality, the clustering will only be based on most eminent components extracted from the SOAP descriptor, which typically are connected to the immediate neighboring shell. With this variability in the fuzziness, our generic algorithm is also geared to a later inclusion into larger workflows, where the resolution as defined by MDS dimensionality and MSC bandwith could for instance be adjusted in active learning cycles evaluating the suitability of the determined equivalence classes for the targeted application.

The following Section \ref{sec:method} will discuss details of the technical implementation of our approach. In Section \ref{sec:result} the performance of the algorithm will be demonstrated by applying it to both finite molecules (Section \ref{subsec:carbon}) and extended materials surfaces (Section \ref{subsec:Pd}). On the basis of these results, we will then discuss limitations and possible extensions in Section \ref{subsec:discussion}.

\section{Methods}\label{sec:method}

\subsection{Environment Representation through SOAP}\label{subsec:SOAP}

\begin{figure}[h]
\centering
\includegraphics[width=0.75\columnwidth]{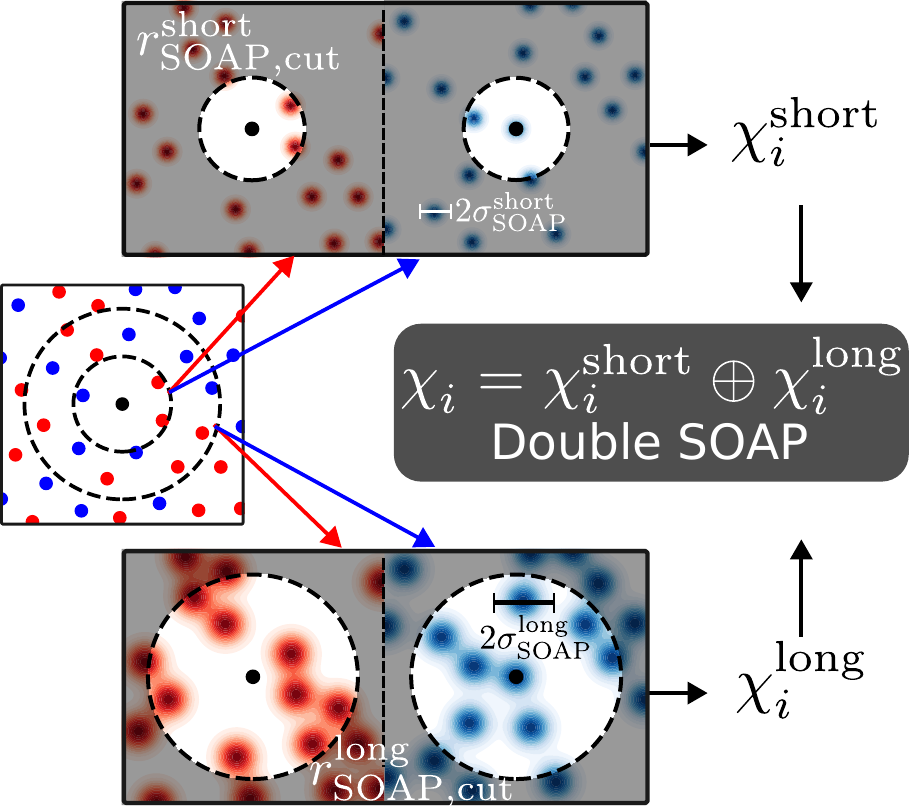}
\caption{Illustration of the double SOAP representation of the local atomic environment of a bonded structure consisting of two chemical species A (red) and B (blue). The panel on the left shows the locations of the atoms in the vicinity of the atom $i$ (black central dot) for which the local environment is to be mapped. The upper and lower row illustrates the short- and long-range part of the double SOAP approach, respectively, see text. For each part, a Gaussian density of width $\sigma^{\rm short/long}_{\rm SOAP}$ is placed at the position of each atom. Summing the densities of the same species gives a total overlapped atomic density for each species separately. Within the corresponding cut-off radius $r_{\rm SOAP, cut}^{\rm short/long}$, these densities are expanded into basis functions. The SOAP vector $\mathbf{\chi}_{i}^{\rm short/long}$ is constructed from the power spectra of the coefficients of these expansions. The double SOAP vector $\mathbf{\chi}_{i}$ of atom $i$ is finally formed by concatenating the short- and long-range SOAP vectors.}\label{fig:SOAP}
\end{figure}

In particular within the booming field of ML interatomic potentials, much progress has recently been achieved in developing general representations of atomic environments that go beyond a mere recognition of (generalized) coordination numbers \cite{rev_with_descriptor1,rev_with_descriptor2}. By construction, these representations encode for instance fundamental symmetries like translational and rotational symmetry, as well as symmetry with respect to permutation of atoms of the same species. Among these representations, we choose for this work the vectorial SOAP descriptor, which for the present purposes offers a good compromise between flexibility and ease of use. The developed framework does not depend on this choice though and any other environment descriptor, e.g. a graph-based one, could equally be employed.

Referring to the original literature for details \cite{SOAP_Bartok_2013}, Fig.~\ref{fig:SOAP} provides an illustration of the working principle of SOAP. In short, it places a Gaussian density function with variance $\sigma_{\rm SOAP}^2$ at the location of each atom within a sphere with radius $r_\text{SOAP,cut}$ centered around the atom $i$ for which the local environment shall be mapped. The overlapped local density of each chemical species is then expanded into a product basis of spherical harmonics for the angular dependence, and a set of orthogonal basis functions for the radial dependence. At this point, one set of expansion coefficients is obtained for each chemical species. To achieve rotational invariance, a normalized power spectrum is subsequently constructed between all combinatorial pairs of coefficient sets of the involved chemical species. This power spectrum is an abstract vector ${\bf \chi}_{i} \in {\mathbb{R}}^S$ describing the local environment like a fingerprint. The dimensionality $S$ of the vector is thereby determined by the number of chemical species and the parameters for the SOAP expansion, namely the maximum $n_\text{SOAP,max}$ and $l_\text{SOAP,max}$ for the radial and angular basis functions, respectively.

Here, we specifically use the so-called double SOAP approach {\cite{double_soap_1,double_soap_2}}, which distinguishes two spheres around the central atom. Higher values for $n^\text{short}_\text{SOAP,max}$ and $l^\text{short}_\text{SOAP,max}$ are chosen in a smaller sphere with radius $r^\text{short}_\text{SOAP,cut}$, while neighboring atoms lying beyond $r^\text{short}_\text{SOAP,cut}$ but within a sphere of radius $r^\text{long}_\text{SOAP,cut}$ are less resolved with lower values $n^\text{long}_\text{SOAP,max}$ and $l^\text{long}_\text{SOAP,max}$. This way, the principle of nearsightedness is naturally built into the environment representation, placing less weight on more distant atoms in the outer sphere, and completely neglecting any neighboring atoms beyond $r^\text{long}_\text{SOAP,cut}$. The two SOAP vectors of the two spheres are then concatenated to form the final SOAP vector $\mathbf{\chi}_i$. We will specify the SOAP parameters used in the different examples in Section \ref{sec:result} below. The package DScribe \cite{Dscribe} is used for the SOAP vector generation throughout this work.

\subsection{MDS Dimensionality Reduction for Clustering}\label{subsec:MDS}

The dimensionality $S$ of double SOAP vectors is very high and easily exceeds several hundreds. This can be a hazard in the clustering process. In particular to also achieve an easily tunable fuzziness, we next map the ${\bf \chi}_i$ first onto a low-dimensional space using MDS \cite{MDS_Carroll_1998}. While the multiple SOAP parameters would therefore in general be chosen to achieve an accurate and non-system specific representation of the local environment, the truly distinctive features of this environment then emerge naturally through this embedding. The dimensionality $S'$ of the corresponding MDS space is thereby a tuning hyperparameter for the fuzziness, which as will be seen below can be as low as two.

MDS is a general technique to map data onto an abstract space while preserving the dissimilarity among data points \cite{MDS_Carroll_1998}. In MDS, dissimilarity is interpreted directly as the distance between data points and in the context of this work, the Euclidean distance between the double SOAP vectors of atoms $i$ and $j$
\begin{equation} 
D_{ij} \;=\; D({\bf \chi}_i, {\bf \chi}_j) = \|{\bf \chi}_i-{\bf \chi}_j\|_2 \quad ,
\end{equation}
is an obvious choice for the dissimilarity of the two local environments. Other kernel forms{\cite{representation_chemrev,structural_comparison}} are conveniently available thanks to the vectorial nature of SOAP. For a total of $N$ atoms in the bonded structure(s) under consideration, this yields a $(N \times N)$ matrix $\bf D$, for which classical MDS solves the eigenvalue problem of the Gram matrix $\bf G$,\cite{MDS_Carroll_1998}
\begin{equation}\label{eq:Gram}
G_{ij} \;=\; \frac{1}{2N}\sum_{k}D_{ik}^2+\frac{1}{2N}\sum_{k}D_{kj}^2-\frac{1}{2N^2}\sum_{k,l}D_{kl}^2-\frac{1}{2}D_{ij}^2 \quad .
\end{equation}
The result is a set of eigenvalues $\lambda_a$ with corresponding normalized orthogonal eigenvectors $\textbf{v}_a = (v_{a,1}, v_{a,2}, \ldots, v_{a,N})^\top$, where $a$ is the index of descendingly ordered $\lambda_a$.

The eigenspace of $\textbf{G}$ can now be used to create a mapping from the SOAP space $\mathcal{S}$ to the abstract embedded space $\mathcal{S}'$. For a chosen dimensionality $S'$ of this MDS space $\mathcal{S}'$, this starts by setting up the $(N \times S')$ matrix $\textbf{P}$ from the first $1 \leq a \leq S'$ eigenvalue-weighted eigenvectors
\begin{equation}\label{eq:mapping}
    P_{ak} \;=\; v_{a,k}/\sqrt{\lambda_{a}} \quad .
\end{equation}
Now consider any atom $i$. The $a$th component of the $S'$-dimensional mapped SOAP vector ${\bf \chi}'_i$ is then \cite{Penrose_Bengio_2003,Penrose_Gower_1968} 
\begin{equation}\label{eq:mapping_op}
    \chi'_{i,a} \;=\; \sum_{k=1}^{N} P_{ak} D^2(\mathbf{\chi} _k,\mathbf{\chi} _i) \quad .
\end{equation}
In other words, Eq.~\eqref{eq:mapping} defines the embedding projector $\textbf{P}$ from the high-dimensional SOAP space $\mathcal{S}$ to the low-dimensional MDS space $\mathcal{S}'$. With the choice of Euclidean distance as the dissimilarity measure, the MDS eigen problem is equivalent to that of a principal component analysis (PCA). This allows us to use the set of eigenvalues $\lambda$ as a guidance selecting a suitable dimension $S'$ e.g. through the broken-stick method{\cite{jolliffe_PCA}}, which does not involve any extra hyperparameter. Specifically, $S'$ is estimated such that for all $a\leq S'$, the normalized eigenvalues are larger than the broken-stick model series $l_a$, $\lambda_a/\sum_k^N\lambda_k \geq l_a$, where $l_a${\cite{jolliffe_PCA}} is,
\begin{equation}\label{eq:broken-stick}
 l_a=\frac{1}{N}\sum_{k=a}^N \frac{1}{k}.
\end{equation}
This approach is also illustrated in Section {\ref{sec:result}}. Alternatively, $S'$ may be seen as a tunable hyperparameter that may e.g. be optimized within a larger workflow that assesses the performance of the classification for a targeted application. Note also, that the embedding operator $\textbf{P}$ works equally for atoms of any additional structure not contained in the original set. This allows to conveniently analyze new structures in terms of a once achieved fuzzy classification, as will be illustrated below. We would like to emphasize that besides classical MDS, there are, of course, other options for dimensionality reduction{\cite{cheng2020SOAP_heuristics}}, as e.g. kernel principal component analysis{\cite{kPCA}} or Sketch Map{\cite{sketch-map,sketch-map2}}. Each of them comes with pros and cons. With other reduction options in general, the number of intrinsic dimensions can also be estimated with other packages e.g. DADApy{\cite{dadapy}}. Here, MDS is our primary choice because of its minimum number of hyperparameters, namely the embedding dimension. Other reduction methods are demonstrated in the SI.

\subsection{Clustering of Atomic Environments}\label{subsec:MSC}

\begin{figure}[h]
\centering
\includegraphics[width=0.5\columnwidth]{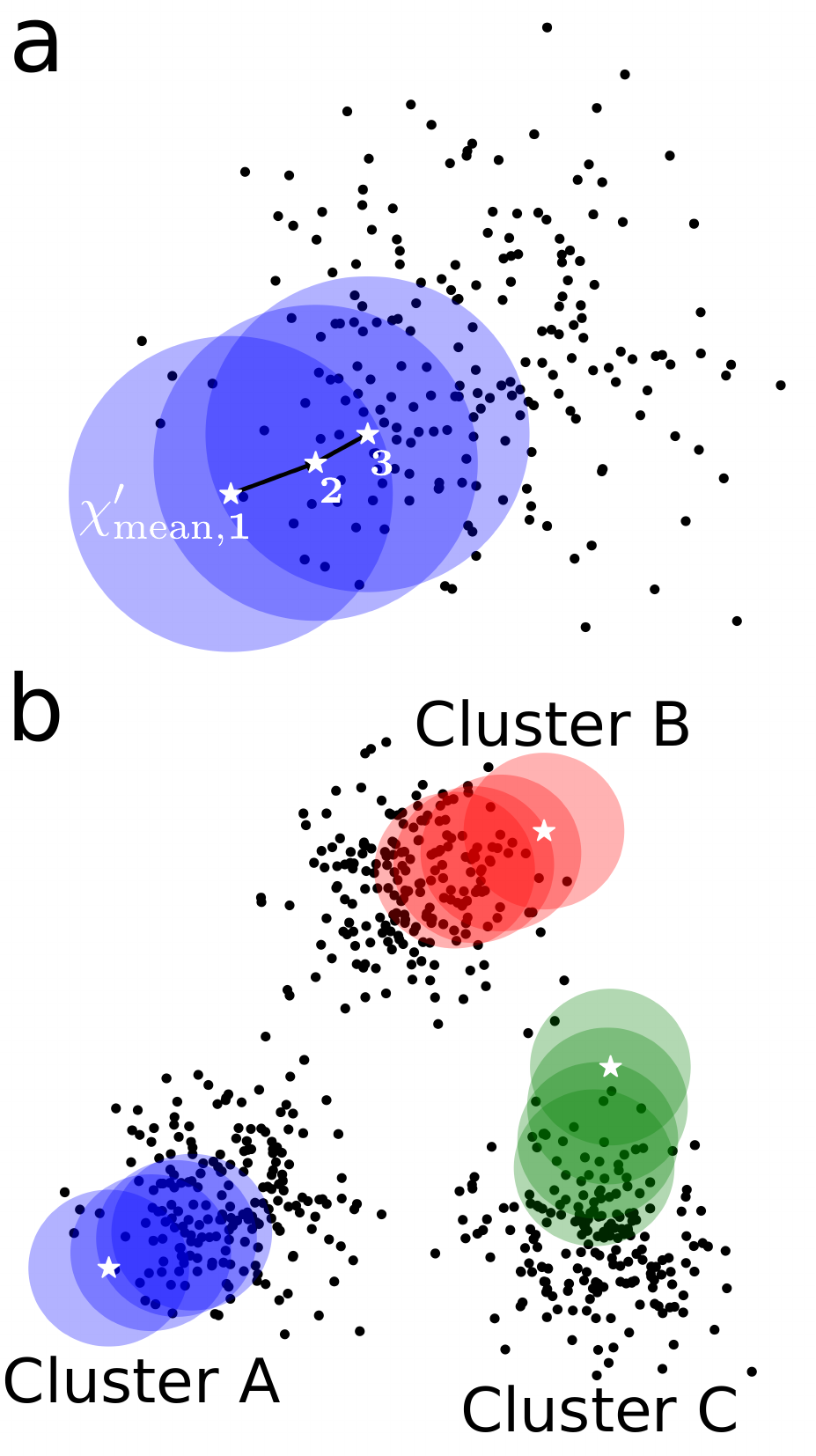}
\caption{a) Illustration of iterations in a typical MSC algorithm. The white stars are the means in the first three iterations, see text. b) Examples of locating three cluster centers by starting the MSC iteration at different data points. The white stars are the starting points of three MSC runs.}\label{fig:MSC}
\end{figure}

Having mapped the dataset to the low-dimensional MDS space $\mathcal{S}'$, we finally cluster the atomic environments according to the geometric similarity reflected in the spatial distribution of the corresponding $N$ data points in $\mathcal{S}'$. This grouping is achieved by mean shift clustering \cite{MeanShift_Fukunaga_1975}, where we employ the implementation of the Scikit-learn package \cite{scikitlearn}. We choose MSC, as it does not require to predefine the final number of groups of equivalent atoms. Instead, its only input parameter is a characteristic distance, the MSC bandwidth $\delta_{\rm MSC}$, which thus emerges as the second tunable hyperparameter of our framework. As with the choice of the SOAP representation before, we note that the developed framework is not restricted to the choice of MSC. As with the choice of the SOAP representation before, we note that the developed framework is not restricted to the choice of MSC. One could well substitute MSC with other density-based cluster algorithms with similar capabilities, as e.g. DBSCAN {\cite{DBSCAN_Ester_1996}}, HDBSCAN{\cite{HDBSCAN,HDBSCAN2}} or spectral clustering {\cite{SpectralClustering_Shi_2000}}. Our current choice of MSC is motivated by its flexibility in handling both noisy and non-noisy datasets and its convenience in out-of-sample classification. For the case of very noisy datasets, HDBSCAN{\cite{HDBSCAN,HDBSCAN2}} might indeed be a more efficient choice. As illustrated in the SI, its performance seems not so good for non-noisy datasets though.

A simple illustration of MSC is shown in Fig. \ref{fig:MSC}a. We start with considering the collection of $N$ data points in $\mathcal{S}'$. A sphere with radius $\delta_{\rm MSC}$ is drawn around any one chosen data point and the mean position $\mathbf{\chi}'_{\text{mean},1}$ of all data points within the sphere is determined. Next, we calculate the mean position $\mathbf{\chi}_{\text{mean},2}$ of all data points within a new sphere around $\mathbf{\chi}'_{\text{mean},1}$ with the same radius $\delta_{\rm MSC}$, cf. Fig. \ref{fig:MSC}a. The mean position is now shifted from $\mathbf{\chi}'_{\text{mean},1}$ to $\mathbf{\chi}'_{\text{mean},2}$. This iteration goes on until the location of the mean converges, which finally gives the center location of a cluster. Starting the algorithm subsequently from all $N$ data points, cf. Fig. \ref{fig:MSC}b, yields a complete list of cluster centers. The number of these clusters is generally lower than $N$ as cluster centers will have coincided during the iterative determination of their location. Each data point is finally assigned to its nearest cluster center.

The bandwidth $\delta_{\rm MSC}$ crucially determines the resolution of the clustering algorithm. With a too large $\delta_{\rm MSC}$, the algorithm will fail to differentiate non-equivalent groups. With a too small $\delta_{\rm MSC}$, it isolates every atom (aka point in the MDS space) into its own group. As with the MDS dimension $S'$, one may simply consider $\delta_{\rm MSC}$ as a tunable hyperparameter of our framework, that could e.g. be optimized by a higher-level workflow into which the present framework is integrated and which evaluates the performance of the achieved fuzzy classification for the targeted application. Alternatively, a simple heuristics for the bandwidth may also be employed. Clusters in the MDS space $\mathcal{S}'$ distinguish themselves by closer distances among their data points than distances to other data points. They thus manifest themselves as agglomerations in the distribution of pairwise distances $D({\bf \chi'}_i, {\bf \chi'}_j) = \|{\bf \chi'}_i-{\bf \chi'}_j\|_2$. For the finite number of $N$ data points, this distribution corresponds to a set of $N(N-1)/2\;\; \delta$-peaks. For larger numbers $N$, identifying distance regions with more or less $\delta$-peaks may then become cumbersome. We therefore conveniently smear every $\delta$-peak into a Gaussian of width $\sigma_{\rm smear}$ and add all Gaussians to arrive at a smooth distribution $D({\bf \chi'}_i, {\bf \chi'}_j) = \|{\bf \chi'}_i-{\bf \chi'}_j\|_2$ that resembles a spectrum. In this spectrum, agglomerations of similar distances will simply show up as peaks. Choosing $\delta_{\rm MSC}$ accordingly somewhere in the minimum after any dominant peak in the smoothed $D({\bf \chi'}_i, {\bf \chi'}_j) = \|{\bf \chi'}_i-{\bf \chi'}_j\|_2$ spectrum should correspondingly yield a good heuristics to identify clusters. A $\delta_{\rm MSC}$ chosen in the first minimum of $D({\bf \chi'}_i, {\bf \chi'}_j)$ will thereby resolve a maximum of clusters, while with a $\delta_{\rm MSC}$ chosen at later minima less and less clusters will be resolved. 

While convenient, the smoothing admittedly adds in principle another empirical parameter $\sigma_{\rm smear}$ to our scheme. In practice, a suitable value for it may readily be found from visual inspection of the smoothened distribution $D({\bf \chi'}_i, {\bf \chi'}_j)$. A more automatized approach recognizes that at any finite temperature vibrations of the atoms in the bonded structures will lead to small changes in the local environment of every atom. Time-averaged, these changes will broaden every point in MDS space and correspondingly every $\delta$-peak in $D({\bf \chi'}_i, {\bf \chi'}_j) = \|{\bf \chi'}_i-{\bf \chi'}_j\|_2$ into a finite Gaussian, too. A useful value for $\sigma_{\rm smear}$ may therefore naturally be determined by analyzing data from molecular dynamics (MD) simulations or when estimating the effect of harmonic displacements on the SOAP vectors. Using a Nose-Hoover thermostat, $NVT$ MD data generated for a large Pd fcc bulk cell at room temperature with $0.5$ fs time steps, e.g. gives the $\sigma_{\rm smear} = 9.20 \times 10^{-3}$ that we employ in the examples below.  The same MD setup was used to generate a $15$ps $NVT$ trajectory for the island on Pd(100) surface structure described below. Equilibration was reached after $2$ps, and 10 snapshots were extracted at random later times to analyze the performance of the framework for the case of finite temperature dynamics.

\section{Results}\label{sec:result}

To demonstrate the versatility of the developed framework we consider two largely different classes of structures. The first, molecular class comprises polycyclic aromatic hydrocarbons (PAHs), while the second class covers various crystalline Pd surfaces. All PAH structures are ideal and generated with nearest-neighbor C-C and C-H distances of 1.42\,{\AA} and 1.08\,{\AA}, respectively. C 1s Kohn-Sham values for these ideal structures were calculated with density-functional theory (DFT) using the FHI-aims package {\cite{FHIaims}} and PBE functional{\cite{PBE}}. The result is presented in the SI. For the generation of the Pd surface we employ an embedded atom potential potential \cite{PdEAM}, which yields a bulk Pd-Pd nearest-neighbor distance of 2.75\,{\AA}. All surface structures are then relaxed until residual forces fall below 0.001\,eV/{\AA}, which already introduces some non-ideality requiring a fuzzy classification. All this data and the entire code used to achieve the fuzzy classifications of the examples discussed in this work can be retrieved from the EDMOND repository. Please refer to the data availability statement for the URL.

As already mentioned above, the purpose of the initial SOAP representation is to provide an accurate and non-system specific description of the local atomic environments, while the truly decisive features governing the fuzzy classification emerge in the subsequent MDS embedding step. As such we simply set all SOAP specific parameters conservatively according to heuristics presently used in the field of ML interatomic potentials (for which SOAP was originally developed) \cite{cheng2020SOAP_heuristics}. Namely this is $n^{\rm short}_\text{SOAP,max} = 8, l^{\rm short}_\text{SOAP,max} = 4, n^{\rm long}_\text{SOAP,max} = 4,$ and $l^{\rm long}_\text{SOAP,max}= 3$. $r^{\text{short}}_{\rm SOAP, cut}$ is conveniently set to a value that corresponds to the mean between the first and the second coordination shell distance in a representative structure for the considered class, whereas $r^{\text{long}}_{\rm SOAP, cut}$ is set at the middle of the third and fourth coordination shell. Here, the representative structures for the two classes are graphene and bulk fcc Pd, which then leads to $r^{\text{short}}_{\rm SOAP, cut}=1.940$\,{\AA} and $r^{\text{long}}_{\rm SOAP, cut} = 3.550$\,{\AA} for the PAHs, and $r^{\text{short}}_{\rm SOAP, cut}=3.320$\,{\AA} and $r^{\text{long}}_{\rm SOAP, cut} = 5.132$\,{\AA} for the Pd surfaces. The Gaussian width $\sigma^{\rm short/long}_{\rm SOAP} = {r^{\rm short/long}_{\rm SOAP, cut}} /8$ in the density representation is chosen proportional to the corresponding cutoff.

For the present illustration purposes, these SOAP settings are fully sufficient, and neither the SOAP determination step, nor the entire algorithm imposes any significant computational burden. The latter could only start to change for excessively large structural databases with a huge total number $N$ of atoms, or if the classification needs to be repeated at very high frequencies. In such cases, optimizing the SOAP parameters would, of course, decrease the computational effort - at the risk of eventually leading to a too coarse initial representation of the environments. In principle, the SOAP settings may also be optimized for the classification task, e.g. following ideas of Barnard et al.{\cite{SOAP_GAS}}. For the present case studies, the successful classification achieved shows though that the general and simple heuristic settings provide a sufficient initial representation that is also not computationally demanding. The performance with different SOAP settings is further illustrated in the SI.

\subsection{Polycyclic Aromatic Hydrocarbons}\label{subsec:carbon}

\begin{figure}[h]
\centering
\includegraphics[width=0.75\columnwidth]{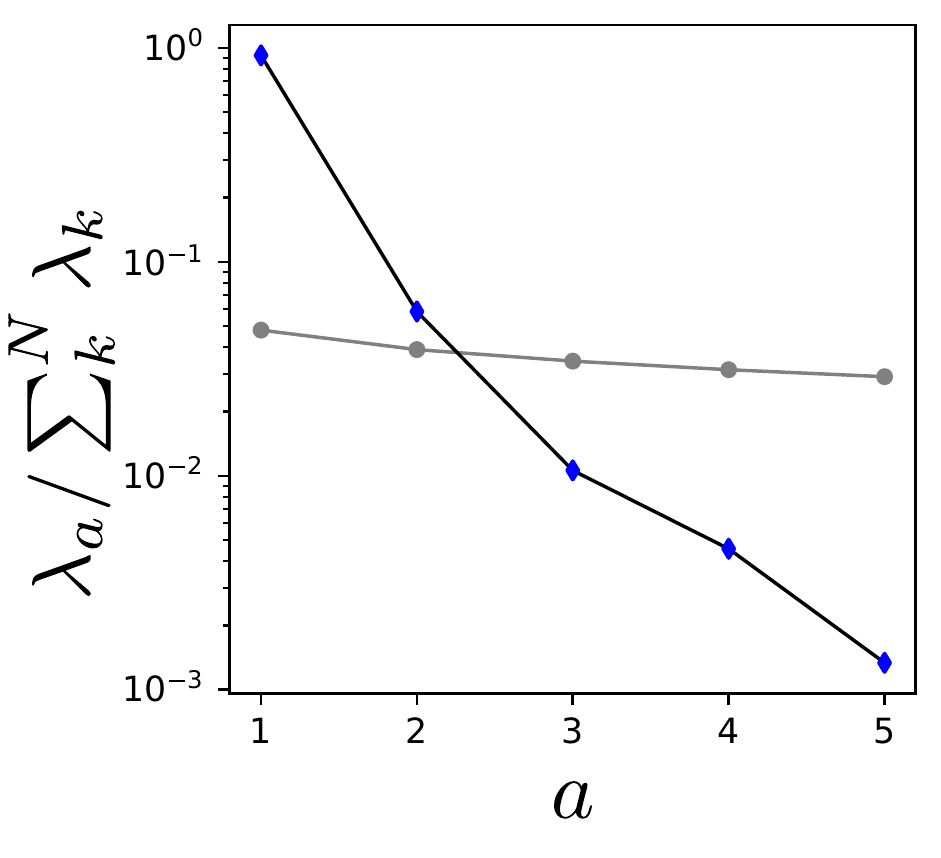}
\caption{The black line shows the normalized eigenvalues $\lambda_a/\sum_k^N \lambda_k$ of the Gram matrix in descending order, cf. Eq.{~\eqref{eq:Gram}}, for the PAH structure set shown in Fig.{~\ref{fig:carbon}}b below. The grey line shows the broken-stick series. Following Eq.{~\ref{eq:broken-stick}}, the estimated suitable MDS dimension $S'$ is 2.}\label{fig:carbon_eigen}
\end{figure}

\begin{figure}[h]
\centering
\includegraphics[width=0.8\columnwidth]{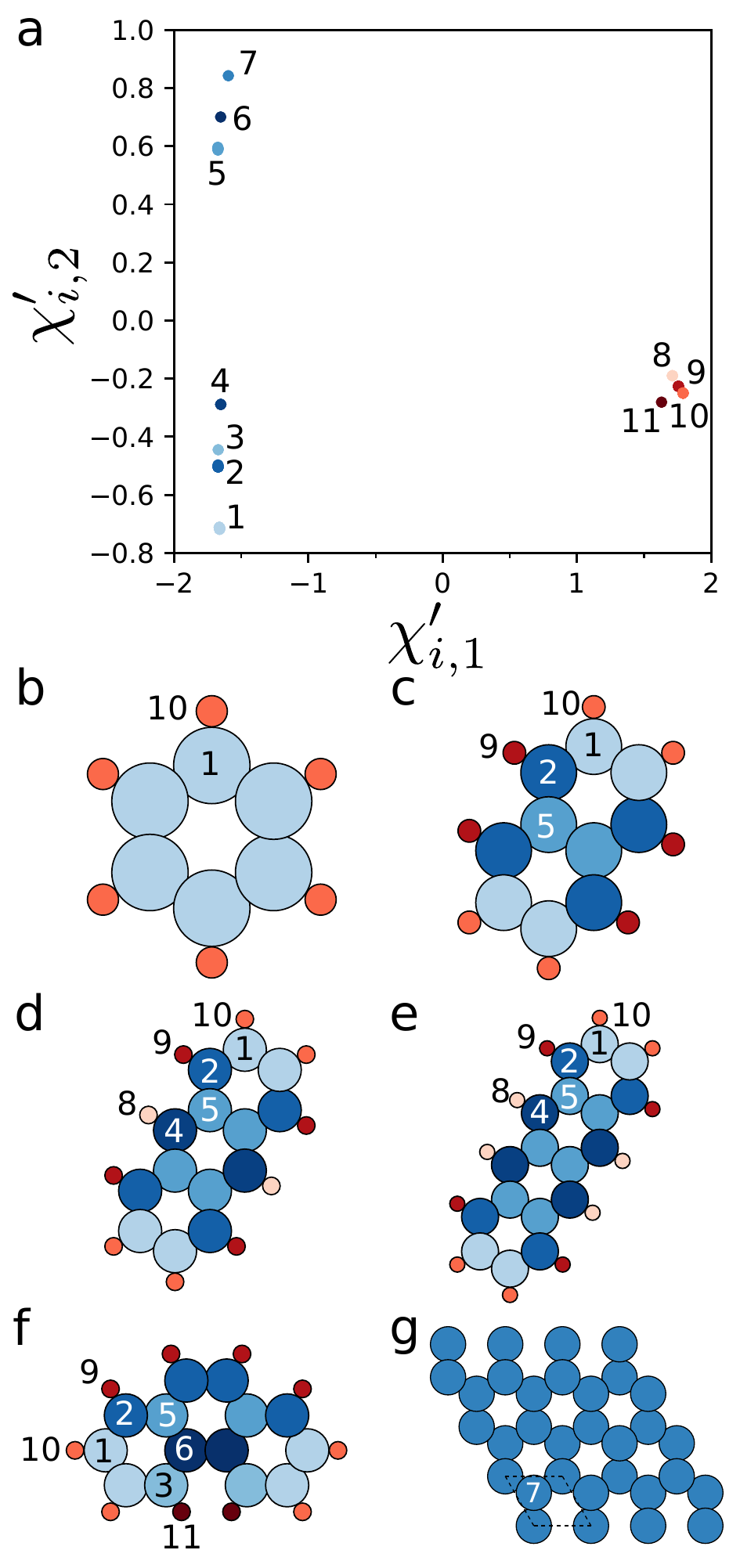}
\caption{(a) Two-dimensional embedded MDS space, in which the $N = 110$ atomic environments contained in the PAH structures collapse into 11 visually separable classes. (b) Structure models of the PAH set, with C atoms drawn as larger spheres and H atoms as smaller spheres. Each atom is colored according to the 11 equivalence classes in (a). For clarity the corresponding class index is shown only once in each structure.}\label{fig:carbon}
\end{figure}

Besides the SOAP representation settings, there are only two relevant hyperparameters left in the framework, both of which tune the resolution of the final fuzzy classification, the MDS dimension $S'$ and the MSC bandwidth $\delta_{\rm MSC}$. As will be seen below, the application to ideal PAH structures renders the determination of the MSC bandwidth trivial and thus provides a good starting point to illustrate the effect of the MDS dimensionality reduction step. The specific PAH set considered is depicted in Fig.~\ref{fig:carbon}b below. It comprises benzene, naphthalene, anthracene, tetracene, phenanthren and a graphene sheet, with a total of $N = 110$ C and H atoms. The normalized eigenvalues of the Gram matrix for this set, cf. Eq.{~\eqref{eq:Gram}}, are shown in Fig.{~\ref{fig:carbon_eigen}}. Following Eq.{~\ref{eq:broken-stick}}, the estimated suitable MDS dimension $S'$ is 2.

Figure~\ref{fig:carbon}a displays the 110 environments embedded in this two-dimensional space. Because of the ideal structures employed, they collapse into 11 visually easily distinguishable classes. The same classification would also result for a wide range of $\delta_{\rm smear}$ for the smoothed pair distance distribution $D({\bf \chi'}_i, {\bf \chi'}_j)$ and the described heuristics to choose the value 0.0241 for the MSC bandwidth somewhere in the first minimum. In the PAH structures shown in Fig.~\ref{fig:carbon}b all atoms are colored according to the thus identified 11 equivalence classes. The automatized algorithm perfectly distinguishes the species and their direct coordinations, just as any human researcher would have done. 

Groups 1-4 are carbon atoms with two carbon neighbors, and groups 5-7 are carbon atoms which have three carbon neighbors; groups 8-11 correspond to hydrogen atoms which all have one neighboring carbon atom.  Obviously, any large enough choice of $\delta_{\rm MSC}$ would have resulted in a clustering that only distinguishes these dominant direct C coordination differences between the three super-groups. However, in the finer resolution of eleven classes, also differences in the arrangement of more distant neighboring atoms are captured.

Which resolution in the classification is more suitable depends on the intended application. We illustrate this in the SI with computed C1s Kohn-Sham levels for these molecules as a simple approximation for core-level spectroscopies {\cite{MichelitschJCP}}. Consistent with the strong difference in C coordination, the levels of each molecule are clustered into two main groups, with the (degenerate) individual peaks reflecting the subtle geometry variations between the C atoms of groups 1-4 and 5-7. It is a question of the experiment, if this substructure in the two main peaks is resolved or not, and corespondingly which clustering bandwidth is more suitably utilized in an automatized computational spectroscopy workflow.

Intriguingly, the differences in the finer resolved 11 classes go beyond mere coordination. For instance, the H atoms in groups 9 and 10 are still resolved even though their local environments only differ in the arrangement of the 2nd neighbor shell. Because of the subtlety of these differences, the corresponding clusters in the MDS space are admittedly very close, cf. Fig.~\ref{fig:carbon}a. Yet, they are still automatically resolved by our framework -- a task that would have been difficult to achieve with predefined symmetry parameters or other classification tools. 

Figure~\ref{fig:carbon}a also nicely demonstrates the added benefit of an increased MDS dimension. The much larger size of the first eigenvalue of the Gram matrix in Fig.~\ref{fig:carbon_eigen} could also have motivated to just choose a one-dimensional MDS space ($S'=1$). Then, the eleven points in Figure~\ref{fig:carbon}a would have all collapsed onto the $\chi'_{i,1}$ axis in Fig.~\ref{fig:carbon}a. At minute distances from each other, the different classes might in principle still have been distinguishable from each other. However, the 2nd embedding dimension separates them much better. The latter would particularly become important, if we consider small deviations from the ideal structures as e.g. induced by thermal vibrations. In that case, the 11 discrete points in Figure~\ref{fig:carbon}a would spread into 11 dense groups of points (or 11 smeared out points if time-averaged MD data is used as discussed above). Then, the MSC clustering would indeed be needed. This is also the case for the Pd surface structure set and we will conveniently discuss the effect of the corresponding $\delta_{\rm MSC}$ hyperparameter for that class in the next section.

\subsection{Crystalline Palladium Surfaces}\label{subsec:Pd}

\begin{figure}[h]
\centering
\includegraphics[width=0.75\columnwidth]{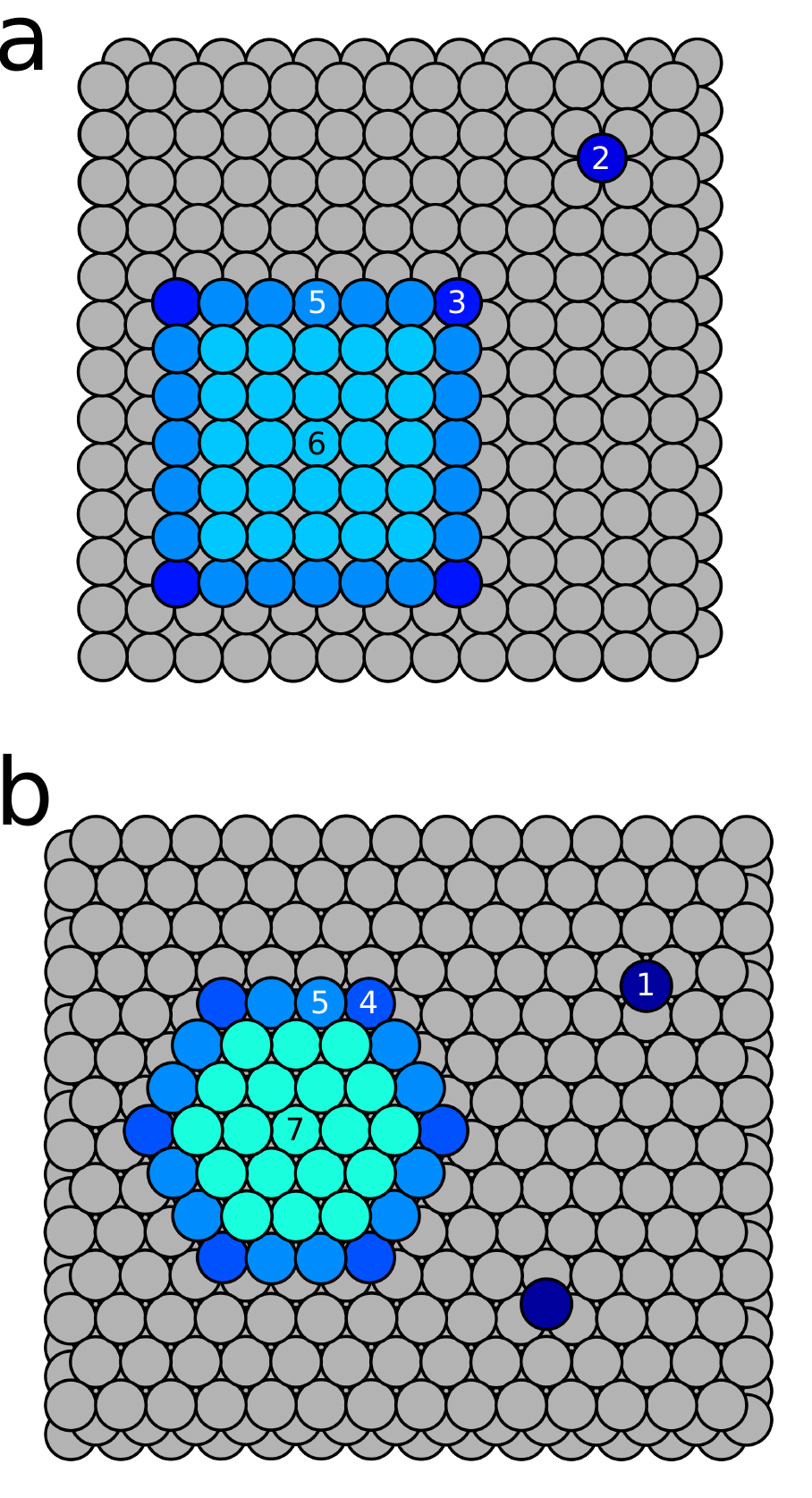}
\caption{Top view of the atomic arrangement of the two nanostructured surfaces contained in the crystalline Pd surface structure set: (a) a $(13\times 13)$-Pd(100) surface with a $(7 \times 7)$ square island and an adatom on top, (b) a $c(14 \times 7\sqrt 3$)-Pd(111) surface with a hexagonal island and two adatoms on fcc and hcp hollow sites. In both cases, groups of atoms discussed in the main text are highlighted with color according to the MSC classification of Fig.~\ref{fig:surface_eigen}b. For clarity, we restrict this coloring to the island atoms and the adatoms, and the corresponding class index is shown only once in each structure.}\label{fig:surface}
\end{figure}

The crystalline Pd surface structure set comprises the low-index Pd(100) and Pd(111) surfaces, as well as the Pd(211) vicinal surface. To represent the extended surfaces, we employ periodic boundary conditions, and for convenience we use in all cases slab geometries as they would also occur in corresponding electronic structure supercell calculations. For the (100) and (111) surfaces we thus use 4 slab layers, and for the (211) surface 5 slab layers. To demonstrate the performance in differentiating surface environments also in more complex cases, we furthermore include two extra structures, namely a $(13\times 13)$-Pd(100) surface with a $(7 \times 7)$ square island and an adatom in a fourfold hollow site; as well as a $c(14 \times 7\sqrt 3$)-Pd(111) surface with a hexagonal island and two adatoms on fcc and hcp hollow sites. Figure~\ref{fig:surface} illustrates the atomic arrangement of these two nanostructures. In total, the Pd surface structure set then contains $N = 1576$ atoms.

\begin{figure}[h]
\centering
\includegraphics[width=0.69\columnwidth]{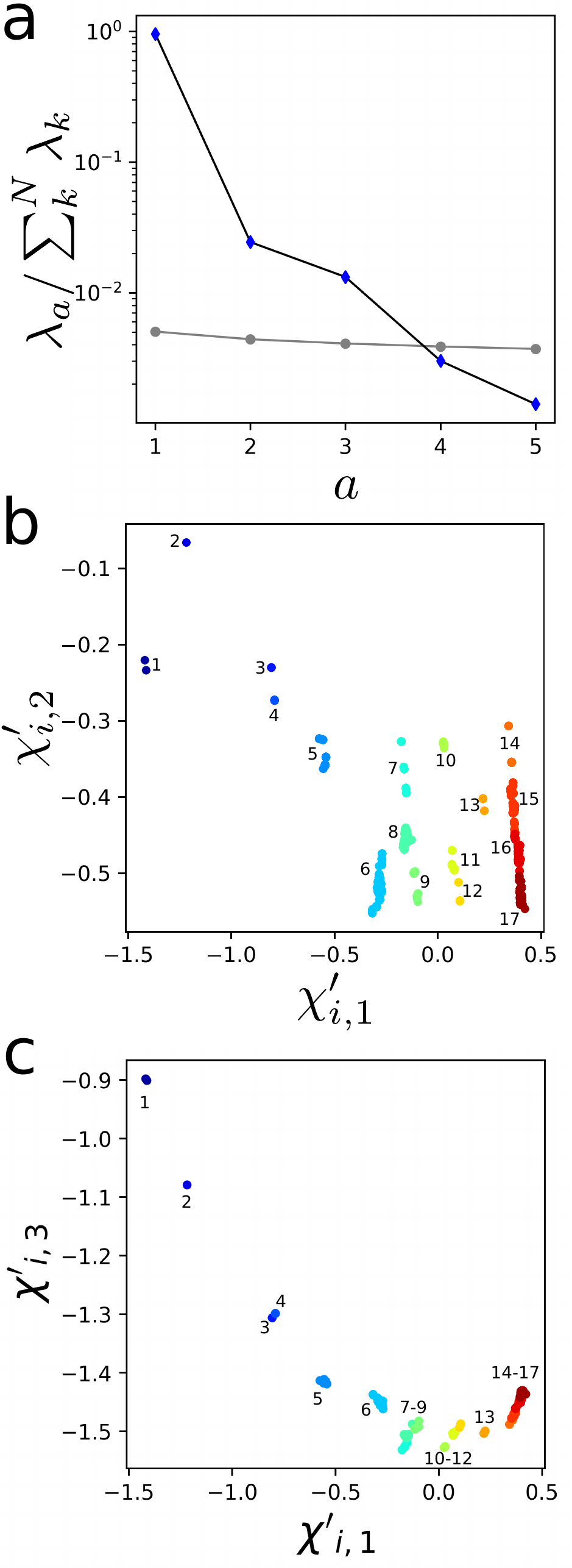}
\caption{(a) The black line shows the normalized eigenvalues $\lambda_a/\sum_k^N \lambda_k$ of the Gram matrix in descending order, cf. Eq.{~\eqref{eq:Gram}} for the crystalline Pd surface structure set shown in Fig.{~\ref{fig:surface}}b. The grey line shows the broken-stick series, following Eq.{~\ref{eq:broken-stick}}, $S'$ is estimated as 3. (b-c) Three-dimensional embedded MDS space, in which the $N = 1576$ atomic environments contained in the Pd surface structures are drawn as individual points. The coloring of the points corresponds to an MSC clustering with a bandwidth of $\delta_{\rm MSC} = 0.0416$ as determined by the simple heuristics, see text. In total 17 equivalent atom classes are identified.}\label{fig:surface_eigen}
\end{figure}
Figure~\ref{fig:surface_eigen}a shows the ordered eigenvalues of the Gram matrix for this set, cf. Eq.~\eqref{eq:Gram}. 
Following Eq.{~\ref{eq:broken-stick}}, $S'$ is estimated as 3. The representation of the $N = 1576$ atomic environments in this space is displayed in Fig.~\ref{fig:surface_eigen}b-c. Not least due to the small geometric differences induced by the surface relaxation, these environments now spread out more than in the ideal PAH example. Yet, they still exhibit a substructure for which even visual inspection suggests some form of clustering. The heuristics to choose the MSC bandwidth in the first minimum of the smoothed distribution of pairwise distances in the MDS space leads to a value $\delta_{\rm MSC} = 0.0416$. The resulting clustering then identifies 17 different classes that are colored and numbered in Fig.~\ref{fig:surface_eigen}b. Analyzing these classes in more detail indicates that the primary MDS dimension $\chi'_{1}$ predominantly distinguishes different coordination numbers, while the other two MDS dimension $\chi'_{2},\chi'_{3}$ seems to more diffusively pick up longer-range arrangement.

The achieved fuzzy classification is demonstrated by color labeling all island surface atoms and adatoms in the two surface nanostructures in Fig.~\ref{fig:surface}. Without relying on any human predefined symmetry parameters, the automatized classification recovers intuitive differences. Adatoms (group 1, 2), island corner atoms (group 3, 4) and island edge atoms (group 5) at the two surface symmetries are correctly distinguished. This performance also extends to the regular surface atoms of the Pd(100), Pd(111) and Pd(211) surfaces, which are all categorized as would be expected from visual inspection. 

On the other hand, one also has to recognize that the resolution achieved by the heuristics is not perfect. This is most straightforwardly seen for the two adatoms on the Pd(111) surface shown in Fig.~\ref{fig:surface}. Both adatoms are classified into the same group 1, even though one of them sits in a hcp hollow site and the other one in an fcc hollow site. At the present MDS dimension and MSC bandwidth, the framework is thus not able to distinguish the differences in the positioning of the 2nd layer Pd atoms between these two sites. The same problem applies to group 5, which contains multiple types of edge atoms. As a result, the color coding of the island on the (111) surface suggests a sixfold symmetry, whereas in reality the symmetry should only be threefold (compare the position of the edge atoms to the underlying Pd terrace atoms).

\subsection{Discussion}\label{subsec:discussion}

On the positive side, the developed framework achieves an automatized fuzzy classification even when resorting to simple heuristics for the determination of its two central hyperparameters, $S'$ and $\delta_{\rm MSC}$. The performance shown for two completely different structural datasets attests to the versatility of the approach, that neither requires any system-specific input, nor predefinition of the number of equivalence classes to be distinguished. On the negative side, already the second, somewhat more involved Pd surface case reveals that the resolution achieved with the heuristic hyperparameters is not perfect. 

Depending on the targeted application, a classification distinguishing even finer details in the local atomic environments might be desirable. As stressed before, we see the main use of the presented fuzzy classification approach as part of a larger workflow, in which active learning loops provide feedback whether the achieved resolution is satisfactory or needs to be increased (or could even be decreased). Take e.g. the initially mentioned application where a starting guess for a transition state search is deduced for an atom by recognizing that it has a similar local environment than another atom for which a transition state is already known. If the efficiency of such guided transition state searches turns out low, this indicates that atoms with too dissimilar environments are fuzzily categorized into the same equivalence class. Provided such feedback, the resolution can then be increased, which in principle should be achievable by increasing the MDS dimension $S'$ and/or decreasing the MSC bandwidth $\delta_\text{MSC}$.

\begin{table}[h]
\begin{center}
\caption{Number of identified equivalence classes for the Pd surface structure set, when systematically increasing the dimension $S'$ of the MDS space, while maintaining the heuristic determination of the MSC bandwidth $\delta_{\rm MSC}$ described in Section \ref{subsec:MSC}. The number of 17 classes resolved for $S' = 3$ was the case discussed in Section \ref{subsec:Pd}.}\label{tab:dimension}
\begin{ruledtabular}
\begin{tabular}{ccc}
$S'$ & $\delta_{\rm MSC}$ & No. of identified classes \\
\hline
1  & 0.0828 &9 \\
2  & 0.0394 &17\\
3  & 0.0416 &17\\
4  & 0.0416 &17\\
5  & 0.0421 &17\\
6  & 0.0425 &18\\
\end{tabular}
\end{ruledtabular}
\end{center}
\end{table}

Unfortunately, there are interdependencies between the two hyperparameters that render a systematic tuning to gradually increase the resolution beyond the one achieved with the heuristic settings difficult. We illustrate this in Table~\ref{tab:dimension} with the number of identified equivalence classes when further and further increasing the MDS dimension $S'$, while maintaining the heuristics-based strategy to determine the MSC bandwidth from the smoothed pairwise distance distribution of the points in MDS space. As expected, the number of resolved equivalence classes does initially increase with larger $S'$. However, it saturates quickly and even in a six-dimensional space the problematic adatom and edge atom cases discussed above are still not properly resolved. The reason for this is that the length scale of the $a$th dimension of the MDS space $\mathcal{S}'$ correlates with the corresponding eigenvalue $\sqrt{\lambda_a}$. Since the $\lambda_a$ are ordered in descending order, the length scales of higher MDS dimensions become smaller and smaller. This can already be seen in the striped structure of the data points in the two-dimensional embeddings in Figs.~\ref{fig:carbon}a and ~\ref{fig:surface_eigen}b. The length scale in the dimension $\chi'_{1}$ is much larger than in the dimension $\chi'_{2}$, and correspondingly the data points are generally more distant from each other in the prior than in the latter dimension. Now, the MSC clustering algorithm determines the mean of the data points within an $S'$-dimensional sphere of radius $\delta_{\rm MSC}$. If the distances between components of the higher MDS dimension become smaller and smaller, adding these dimensions will not help much to further distinguish clusters unless $\delta_{\rm MSC}$ is also reduced. However, as can be seen from Table~\ref{tab:dimension} the simple heuristics to determine $\delta_{\rm MSC}$ from the smoothed pairwise distance distribution instead leads to a roughly constant value for this bandwidth in higher MDS dimensions. Indeed, simply reducing $\delta_{\rm MSC}$ from the presently employed values to below 0.025 will immediately resolve 23 equivalence classes already in $S' = 2$. 

On the other hand, just reducing $\delta_{\rm MSC}$ is neither a general purpose solution. A too small $\delta_{\rm MSC}$ will start to distinguish atoms according to their larger distances in the primary MDS dimensions and maybe such distinction is not desired either. Take the example of the two non-resolved adatoms on the Pd(111) nanostructured surface, i.e. the two points in group 1 in Fig. ~\ref{fig:surface_eigen}b. A sufficiently reduced $\delta_{\rm MSC}$ would allow to distinguish the two. However, at such a small $\delta_{\rm MSC}$ the MSC algorithm will also start to differentiate the numerous bulk-like Pd atoms that currently make up the red stripe at the bottom right in Fig. ~\ref{fig:surface_eigen}b -- and adding further MDS dimensions will not mitigate this problem at all. Alternatively, one could imagine re-scaling the MDS dimensions by their eigenvalue to achieve more comparable length scales in all MDS dimensions. However, the diminishing length scales of higher MDS dimensions have a meaning. They reflect that differences in these dimensions correspond to more and more subtle differences in the local atomic environments. Blowing up these differences by simply renormalizing the MDS length scales might therefore neither be a generally applicable remedy to increase the resolution in a desired way. In the end, a careful tuning of both hyperparameters, $S'$ and $\delta_{\rm MSC}$, will be required, if the default heuristics do not achieve a satisfactory fuzzy classification for a specific application. The classification performance with the heuristics utilizing a different $\sigma_\text{smear}$ is further demonstrated in the SI.

A noteworthy positive feature of the developed framework is that new structures may readily be evaluated within a once achieved fuzzy classification. As long as exactly the same SOAP settings to initially describe the atomic environments are employed, the embedding operator $\textbf{P}$ of this classification will project any environment contained in the new structure to the low-dimensional MDS space $\mathcal{S}'$. The corresponding new data point $\mathbf{\chi}'_\text{new}$ can then straightforwardly be assigned to the nearest cluster center. 

We explore this idea for a room-temperature MD trajectory generated for the Pd(100) island structure. 10 snapshots are extracted at random times and the environments of all surface atoms are categorized in terms of the 17 equivalence classes of the Pd surface structure set discussed above. As summarized in the SI, adatoms, island edge or surface atoms are correctly identified with a 90$\%$ or higher probability despite the thermal displacements. More problematic are only the island corner atoms with their larger anisotropic vibrations, which are miscategorized with a 50$\%$ probability. One option that we currently pursue to improve this could be to exploit correlations in the classification of the individual atoms in successive snapshots along the trajectory.

\begin{figure}[ht]
\centering
\includegraphics[width=1.0\columnwidth]{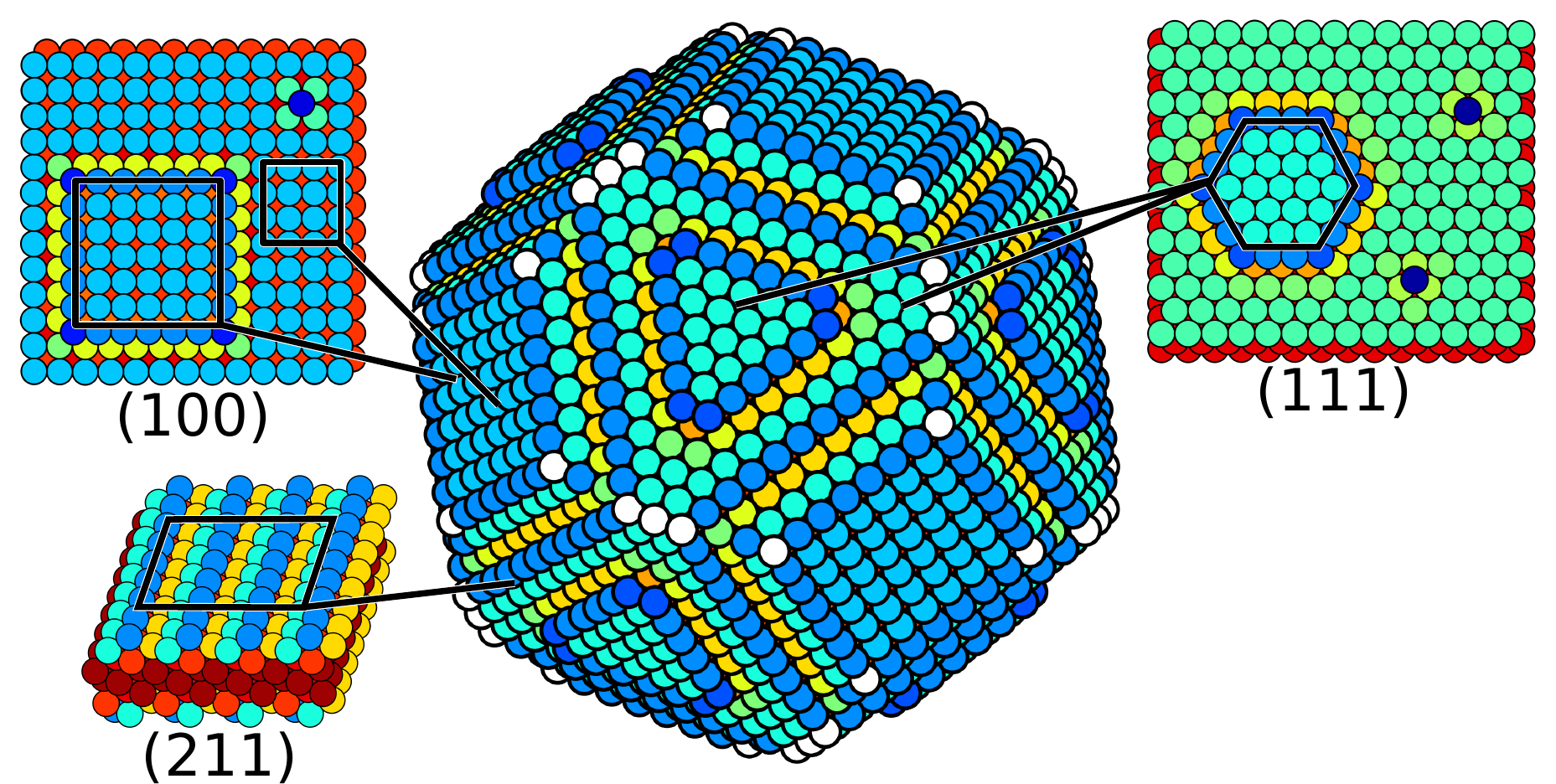}
\caption{Categorization performance for new structures within a once achieved fuzzy classification. The 8523 atoms of the shown Pd nanoparticle are colored according to the 17 equivalence classes of the Pd surface structure discussed above, cf. Fig.~\ref{fig:surface_eigen}b. For comparison, the Pd(211) surface and the two surface nanostructures of this original structure set are also shown in the same coloring, readily allowing to identify similar local environments. White atoms close to facet edges of the Pd nanoparticle are classified as distinct from all classes of the existing classification, see text.}\label{fig:nanoparticle}
\end{figure}

An alternative ansatz, not only for the case of finite temperature dynamics, is to iteratively expand a given fuzzy classification with such new structures. In this case, an atomic environment would be identified as distinctly different from all previously considered environments, if the corresponding new data point {$\mathbb{\chi}'$} is located further away from any existing cluster in MDS space than the MSC bandwidth {$\delta_\text{MSC}$}. In Fig.~{\ref{fig:nanoparticle}}, this is illustrated for the 8523 atoms of a Pd nanoparticle again within the 17 previously discussed equivalence classes. In this case, all edge atoms between the nanoparticle facets are categorized as a new environment. Once one or a sufficient number of such new environments are identified, a flag could be set in an iterative framework that initiates a new fuzzy classification now involving the entire, augmented structure set.
\section{Conclusions}\label{sec5}

We presented an automatized machine-learning framework to identify atoms with (near-)equivalent local atomic environments in any one or a set of given structures. The required fuzziness in the classification is achieved by embedding an initial high-dimensional representation of the local environment and a subsequent clustering in the resulting low-dimensional space. Emphasis was placed on a high versatility of the framework at minimum system specific input. As such the framework is readily applicable to molecular structures, extended materials or interfaces, to ideal or non-ideal, as well as crystalline or amorphous geometries. Simple heuristics are provided for the two central hyperparameters, the dimension $S'$ of the MDS embedding space and the bandwith $\delta_{\rm MSC}$ of the MSC clustering in this space. If the resolution achieved with the heuristic settings is not optimum for a specific application, the two hyperparameters are tunable with understandable effects on the resulting classification. The framework could therefore readily be integrated into larger workflows that achieve an optimum tuning of the hyperparameters e.g. in active-learning iterations evaluating the classification performance for the targeted application. A sample implementation of the framework as a standalone application is provided in the repository stated below.

The versatility of the framework also extends to its capability to assess new structures within an achieved fuzzy classification. Also to this end, one can therefore imagine an iterative usage, in which the atoms of new structures are first assessed and a new fuzzy classification of the increased set of structures is initiated whenever a critical number of new, distinctly different atom classes has been identified. We also note that the initial high-dimensional fingerprint for each atom is not necessarily restricted to the structure-sensitive (double) SOAP vector employed in this work. This representation was chosen here within the focus on equivalence in the local atomic environments. Other fingerprints like partial charges or other electronic structure properties may e.g. be added to further improve the resolving capabilities of the framework, or directly be used to base the fuzzy classification on aspects other than geometric similarity.

\nocite{*}
\section*{Supplementary Material}
Please see the supplementary material for demonstrations of other dimensionality reduction methods (kPCA, Sketch Map); another clustering method (HDBSCAN). Also classification demonstration applied on MD data of structured Pd(100) surface.

\section*{Acknowledgments}
KCL gratefully acknowledges an Alexander von Humboldt Foundation research fellowship.

\section*{Declarations}
\subsection*{Conflict of Interest}
The authors have no conflicts to disclose.
\subsection*{Author Contributions}
\textbf{K.C. Lai}: Conceptualization (equal); Methodology (equal); Data curation (lead); Formal analysis (lead); Investigation (equal); Writing – original draft (equal). 
\textbf{S. Matera}: Conceptualization (equal); Methodology (equal); Investigation (equal); Writing – original draft (equal). 
\textbf{C. Scheurer}: Conceptualization (equal); Methodology (equal); Investigation (equal). 
\textbf{K. Reuter}: Conceptualization (equal); Methodology (equal); Writing – review \& editing (lead); Project administration (lead); Supervision (lead).

\section*{Data Availability Statement}
The data that support the findings of this study are openly available in EDMOND, at DOI:10.17617/3.U7VKBM .

\section*{References}
\bibliography{aipsamp}

\end{document}


\title{Supporting Information to "A Fuzzy Classification Framework to Identify
Equivalent Atoms in Complex Materials and Molecules"}
\author{King Chun Lai}\email{lai@fhi-berlin.mpg.de}
\author{Sebastian Matera}
\author{Christoph Scheurer}
\author{Karsten Reuter}
\affiliation{Fritz-Haber-Institut der Max-Planck-Gesellschaft, Faradayweg 4-6, 14195 Berlin, Germany}

\date{31 May 2023}
             
\renewcommand{\appendixname}{Supplementary Information}
\renewcommand{\thefigure}{S.~\arabic{figure}}
\setcounter{figure}{0}
\renewcommand{\thetable}{S.~\arabic{table}}
\setcounter{table}{0}
\renewcommand{\theequation}{S.~\arabic{equation}}
\setcounter{equation}{0}

\maketitle

\onecolumngrid
\section{Dimensionality Reduction Methods}\label{sec:DR}
In this section, we demonstrate employing other dimensionality reduction methods in the framework. Since our primary choice, classical multidimensional scaling (MDS) with Euclidean distance as dissimilarity measure implies equivalence to principal component analysis (PCA), we will omit any demonstration of PCA. For each method, embedding is applied on the full SOAP vectors $\mathbb{\chi}$ from the Pd dataset in the main text Sec.3B.

\subsection{kernel PCA}\label{subsec:kPCA}
\begin{figure*}[h]
\centering
\includegraphics[width=0.8\textwidth]{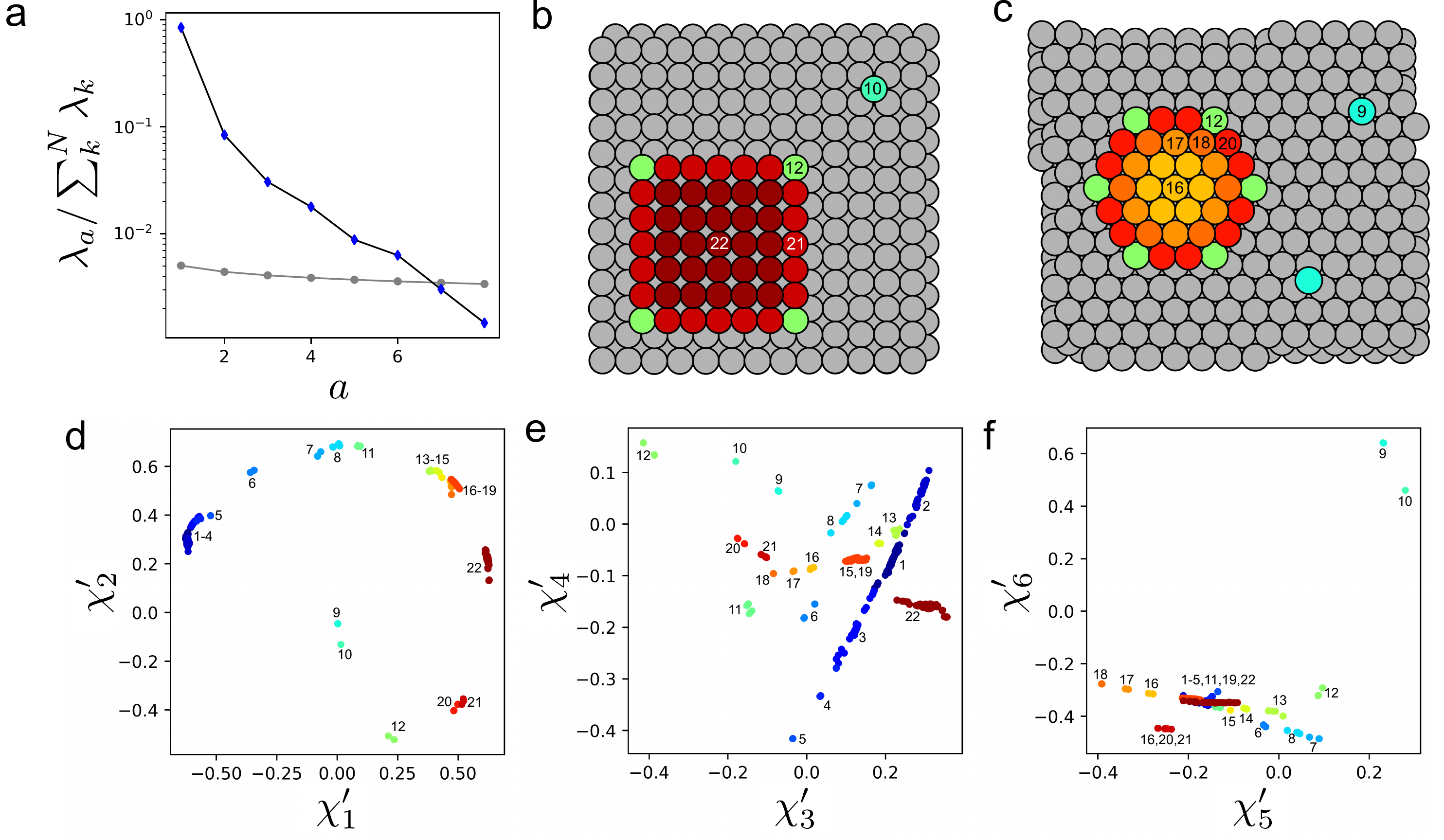}
\caption{kPCA applied to the Pd dataset in the main text Sec.3B. The kPCA hyperparameters are $\sigma_\text{kPCA}=0.137$ and $S'=6$; for the following MSC, the estimated bandwidth is $\delta_\text{MSC}=0.0635$ from the heuristics. (a) The black line shows the normalized eigenvalues of kPCA; the grey line shows the broken-stick-model, leading to a six dimensional embedding. (b-c) Pd surfaces in the dataset with only the island and adatoms highlighted according to (d-f). (d-f) The embedded SOAP vector $\mathbb{\chi}'_i$ with color scheme corresponding to 22 groups.}\label{fig:kPCA_Pd}
\end{figure*}

For kernel PCA (kPCA)\cite{kPCA}, a typical choice of the kernel function $K(\mathbb{\chi}_i, \mathbb{\chi}_j)$ between two SOAP vectors $\mathbb{\chi}_i,\mathbb{\chi}_j$ is a Gaussian kernel in the following form,
\begin{equation}\label{eq:kPCA}
K(\mathbb{\chi}_i, \mathbb{\chi}_j)=\text{exp}\left(-\frac{\|\mathbb{\chi}_i-\mathbb{\chi}_j\|^2}{2\sigma_\text{kPCA}^2}\right),
\end{equation}
with $\sigma_\text{kPCA}$ being a hyperparameter, besides the embedding dimension $S'$ like that in PCA or classical MDS. In this demonstration, we estimate this $\sigma_\text{kPCA}$ by simply rescaling the typical length scale $\sigma_\text{smear}=9.20\times10^{-3}$ used in the main text. Specifically,
\begin{equation}\label{eq:kPCA_sigma}
\sigma_\text{kPCA}=\sqrt{S-1}\sigma_\text{smear},
\end{equation}
with $S=220$ being the dimension of a full SOAP vector $\mathbb{\chi}$, giving $\sigma_\text{kPCA}=0.137$. From Fig.~\ref{fig:kPCA_Pd}a, the broken-stick approach suggests $S'=6$. The result of clustering is shown in Fig.~\ref{fig:kPCA_Pd} with again $\sigma_\text{smear}=9.20\times 10^{-3}$ giving $\delta_\text{MSC}=0.0635$ according to the heuristics.

The result is compatible with the classical MDS in the main text. Compared to classical MDS, edge atoms of islands on (100) and (111) are differentiated (group 20, 21), also various inner (111) island atoms are differentiated (group 16-18). However, the two different types of corner atoms on (100) and (111) are merged into the same group 12. Overall, the heuristics gives reasonable initial hyperparameters under an alternate choice of embedding methods.

\subsection{Sketch Map}\label{subsec:SM}
\begin{figure*}[h]
\centering
\includegraphics[width=0.8\textwidth]{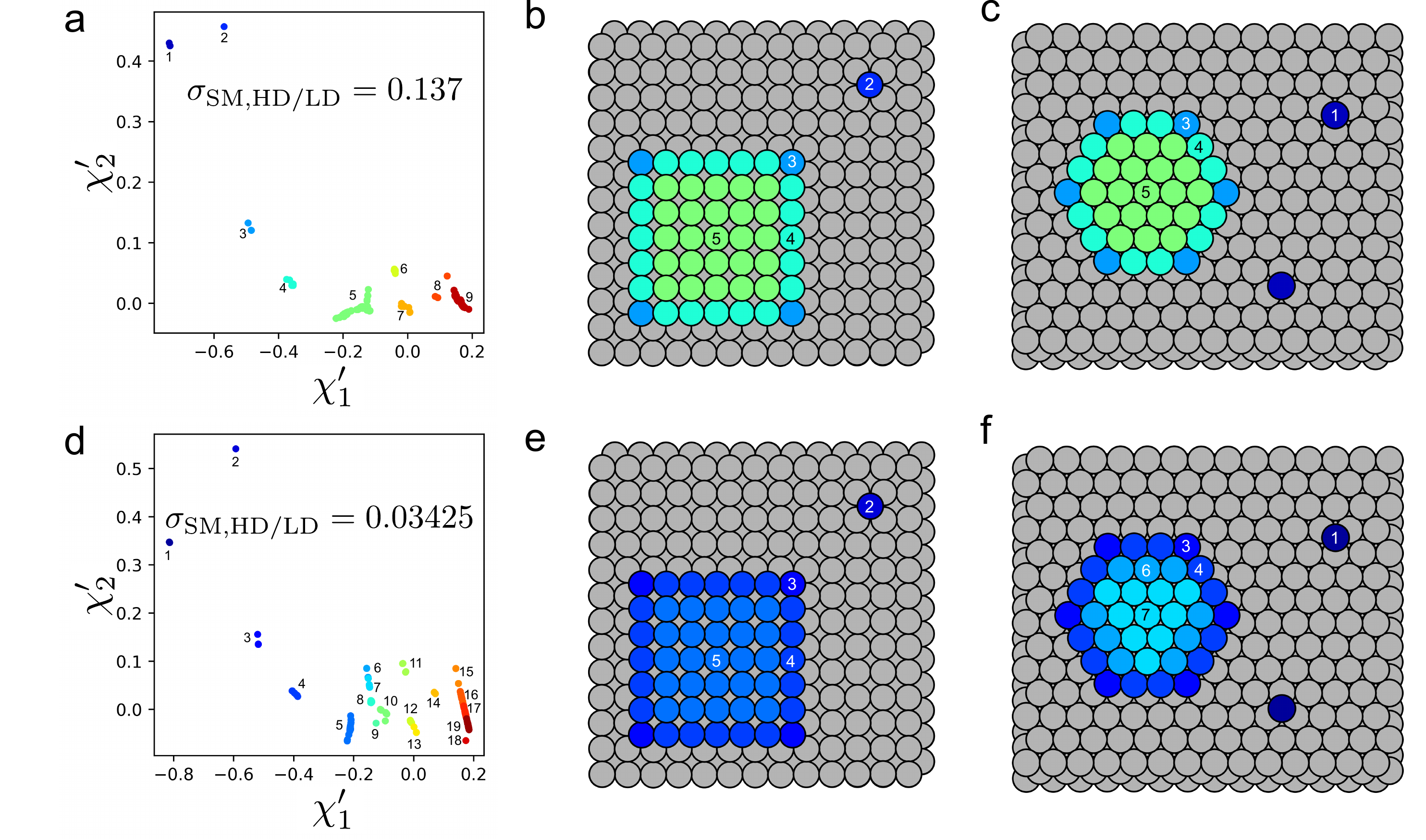}
\caption{(a-c) Sketch Map applied on the Pd dataset in the main text Sec.3B. The sketch-map hyperparameters are $A=S=220$, $B=1$, $a=S'=2$, $b=1, \sigma_\text{SM, HD}=\sigma_\text{SM, LD}=0.137$; for the followed MSC, $\delta_\text{MSC}=0.0377$ is obtained from the heuristics. a) The embedded SOAP vector $\mathbb{\chi}'_i$ with a color scheme corresponds to 9 groups. (b-c) Pd surfaces in the dataset with only the island and adatoms highlighted according to (a). (d-f) Another attempt with $\sigma_\text{SM, HD}=\sigma_\text{SM, LD}=0.03425$, leading to $\delta_\text{MSC}=0.0202$, then 19 groups in MSC. The group indices in these two attempts do not correspond to each other.}\label{fig:SM_Pd}
\end{figure*}

Another recently developed method based on iterative MDS is Sketch Map\cite{sketch-map,sketch-map2}. This introduces a stress function with six major hyperparameters, three for the high-dimension space $\sigma_\text{SM, HD}, A, B$ and three for the low-dimension space $\sigma_\text{SM, LD}, a, b$. Please refer to the original work\cite{sketch-map,sketch-map2} for details. In this demonstration, we will perform a two dimensional embedding for convenience. Following the guidelines\cite{sketch-map2}, we set $A=S=220$, $B=1$, $a=S'=2$, $b=1$. The two $\sigma_\text{SM}$ are chosen to be the same, $\sigma_\text{SM, HD}=\sigma_\text{SM, LD}=0.137$, following the same idea when rescaling $\sigma_\text{kPCA}$ in the kPCA demonstration Eq.~\ref{eq:kPCA_sigma}. To reduce human prior-knowledge about the dataset, we will not pre-select any set of "landmarks" for the embedding. The result is shown in Fig.~\ref{fig:SM_Pd}a-c. With the 8- and 9-coordinated environments being merged into the same group 5, this classification is more coarse-grained than just a coordination number.

Rather than an automated classification, a more detailed tuning in hyperparameters is needed. Given that Sketch Map embedding is sensitive to the choice of $\sigma_\text{SM, HD/LD}$, another attempt with $\sigma_\text{SM, HD/LD}$ being manually set to a quarter of the previous one, $\sigma_\text{SM}=0.00343$ is shown in Fig.~\ref{fig:SM_Pd}d-f. This setting leads to an embedding closer to MDS. For the clustering, it leads to $\delta_\text{MSC}=0.0202$ from heuristics, giving 19 groups. Overall it may take an intermediate embedding to inspire a choice of $\sigma_\text{SM, HD/LD}$ before applying Sketch Map for the best performance as it was demonstrated in the original work\cite{sketch-map2}.

\clearpage

\section{Clustering}\label{sec:Clustering Methods}
There is a wide spectrum of clustering methodologies handling datasets in various approaches, each of them comes with pros and cons. There are two criteria for being a good candidate for the classification framework in this contribution. The first one is being free from a predetermined number of clusters, since obviously the number of environment groups is generally unknown. The second criteria is a well implemented out-of-sample cluster prediction, for an expanding database in machine learning cycles. Unfortunately, there is only a limited number of algorithms satisfying both criteria. Mean shift clustering is our primary choice for its transparency in the meaning of the hyperparameter (bandwidth $\delta_\text{MSC}$), besides satisfying the mentioned criteria. Here, we will demonstrate another clustering method, Hierarchical Density-Based Spatial Clustering of Applications with Noise (HDBSCAN\cite{HDBSCAN}) applied on the MDS-embedded Pd dataset $\mathbb{\chi}'$ used in the main text section 3B.

\subsection{HDBSCAN}\label{subsec:HDBSCAN}
\begin{figure*}[h]
\centering
\includegraphics[width=0.60\textwidth]{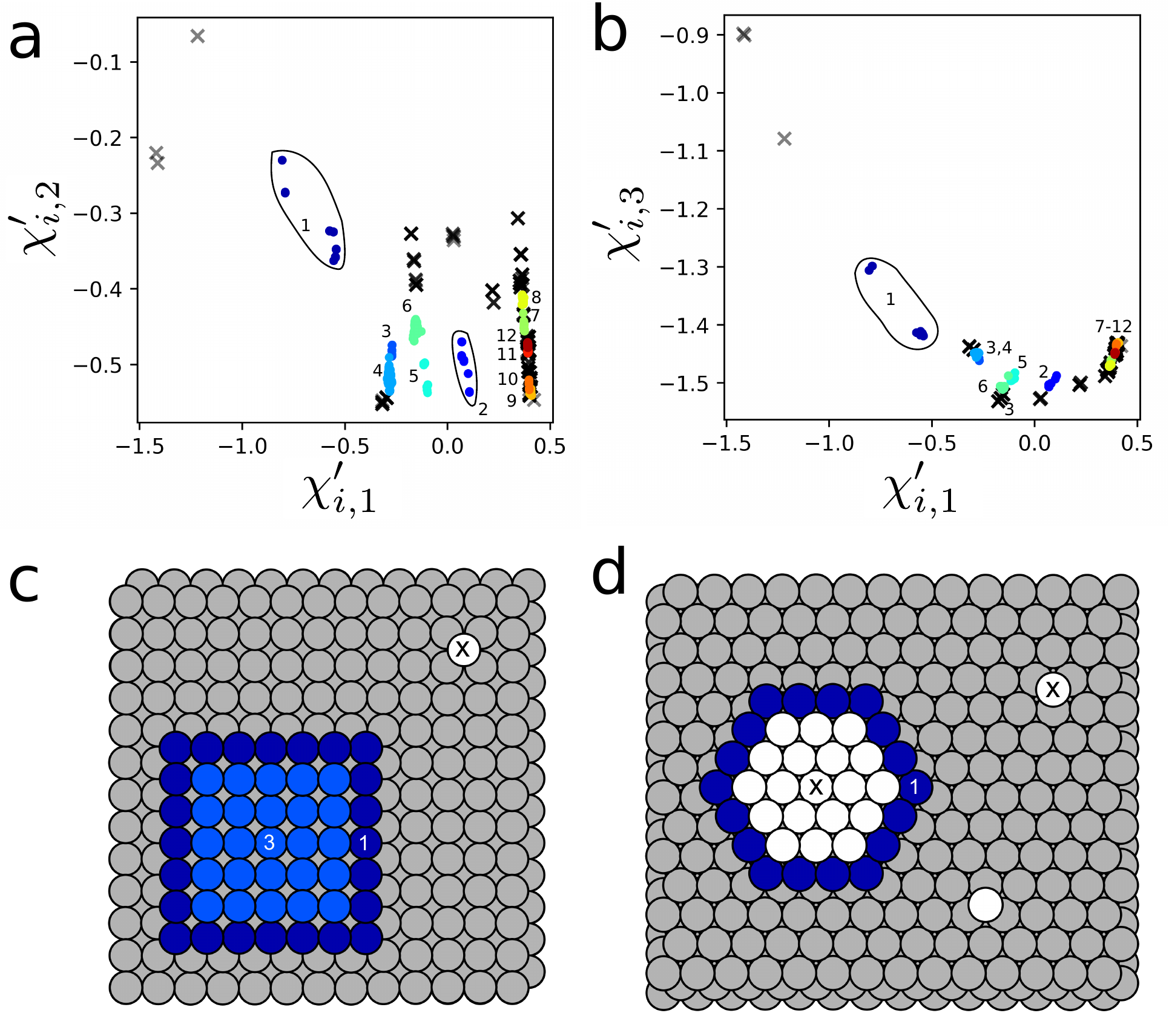}
\caption{HDBSCAN applied on the Pd dataset shown in Sec.3B with hyperparameters $m_\text{pts}=m_\text{clSize}=20$. (a-b) The three-dimensionally embedded SOAP vectors $\mathbb{\chi}'_i$. Out of the 1576 data points, 1360 are categorized into 12 groups, each group is highlighted with one color. The crosses represent 216 data points categorized as noise by HDBSCAN. Group 1 and 2 are circled as they span much wider than other groups. (c-d) Pd surfaces with the islands and adatoms colored according to the groups shown in (a-b), white atoms are categorized as noise.}\label{fig:HDBSCAN_Pd}
\end{figure*}

The advantage of HDBSCAN\cite{HDBSCAN,HDBSCAN2} is its robustness against noisy datasets as its name suggests. However, the disadvantage is a lack of transparency in the impact of these hyperparameters. These hyperparameters are optimized by either scanning the hyperparameter space or being provided manually based on prior-knowledge. There are practically two hyperparameters: the minimal number of data points $m_\text{clSize}$ in a cluster; and the number of neighboring data points $m_\text{pts}$ when the 'core distance' is computed. Please refer the original algorithm\cite{HDBSCAN2} for details. In general speaking, increasing $m_\text{clSize}$ reduces the number of groups but more data points are considered as noise since they fail to be counted as a cluster. On the other hand, reducing $m_\text{pts}$ makes the clustering more conservative and less data points are considered as noise. In the scenario of our embedded Pd dataset with clusters of very different sizes, there is no ideal intuitive choice. We will start with $m_\text{clSize}=20$, and set $m_\text{pts}=m_\text{clSize}$ as that is the default setting. The result is shown in Fig.~\ref{fig:HDBSCAN_Pd}.

\begin{figure*}[h]
\centering
\includegraphics[width=0.3\textwidth]{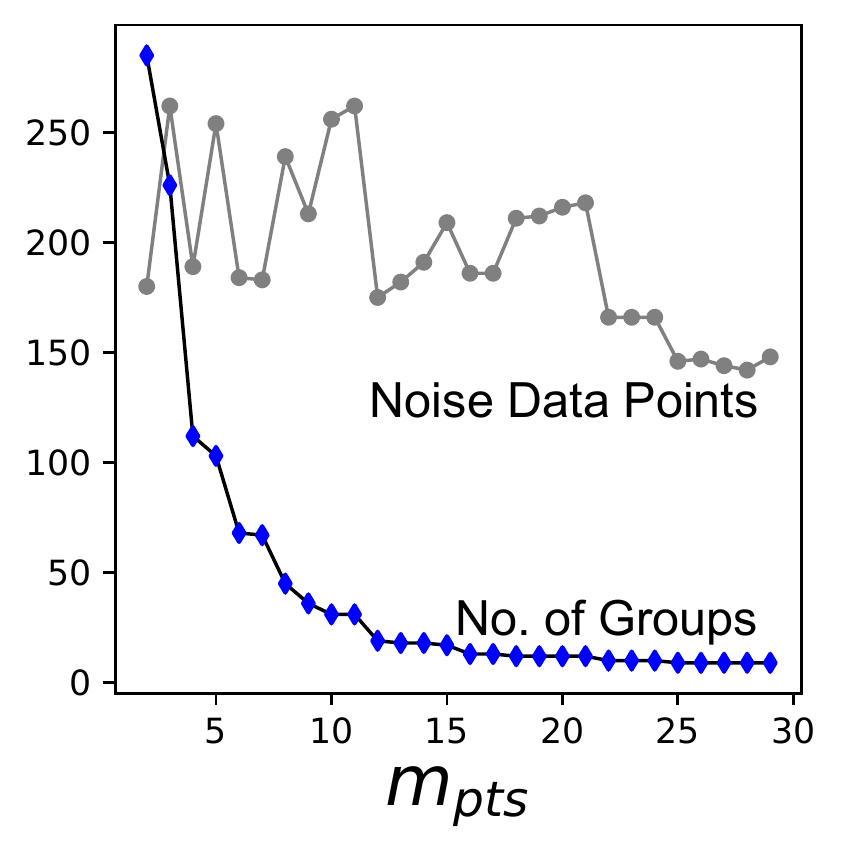}
\caption{Fine tuning the hyperparameters with $m_\text{pts}=m_\text{clSize}$ in a range of 2 to 30. The number of groups is shown in blue diamond points and the number of noise data points are shown in grey dots. Despite the general trend of a reduced number of groups with an increasing $m_\text{pts}=m_\text{clSize}$, the impact on the number of noise data points is not clear.}\label{fig:HDBSCAN_Pd_tune}
\end{figure*}

Out of the 1576 data points, 216 are categorized as noise in this setting. The other 1360 data points are categorized into 12 groups. The environments with very few data points (e.g. adatoms on (100) or (111) surfaces, the three data points at the top left in Fig.~\ref{fig:HDBSCAN_Pd}a) are recognized as noise. Similarly, group 1 spans a wide area and includes both edge and corner atoms, likely due to the low population and thus low density around that part of the MDS space. We by no means imply our choice of hyperparameters are optimized, one might tune these hyperparameters for a better performance. To further explore the effect of these hyperparameters, we scan the values of $m_\text{clSize}\in \left[2,30\right]$ while keeping $m_\text{pts}=m_\text{clSize}$. The number of noise data points and number of groups are shown in Fig.~\ref{fig:HDBSCAN_Pd_tune}. Despite the general trend of a reduced number of groups with an increasing $m_\text{clSize}$, the impact on the number of noise data points is not obvious. Overall, HDBSCAN is not designed for non-noisy datasets, in which isolated data points are not noise but represent independent groups. On the other hand, HDBSCAN can be a more suitable option other than MSC when handling noisy datasets with a comparable population density of each group.

\section{Correlation between clustering and physical properties}\label{sec:CLS}
The SOAP descriptor relies completely on the geometrical information describing the local atomic arrangement. Yet, it is obvious that such configurational arrangement is a key factor impacting local physical quantities. In this section, we demonstrate the correlation between our geometry-based clustering and physical quantities.

\begin{figure*}[h]
\centering
\includegraphics[width=0.8\textwidth]{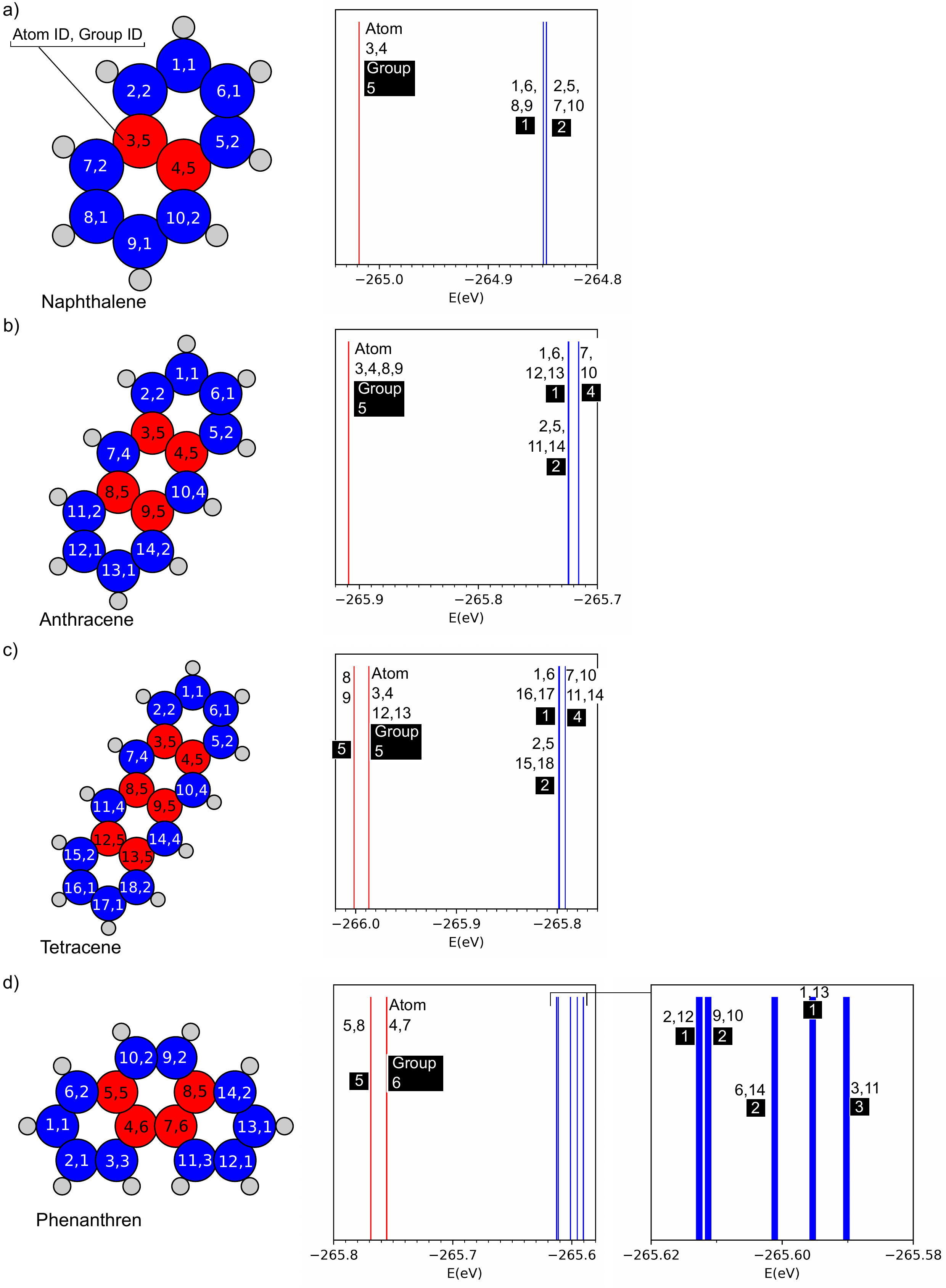}
\caption{(a-d) Each subfigure shows on the left hand side the ideal PAH structure identical to that in the main text, Fig.4c-f. Blue (red) color labels the carbon atoms with C coordination 2 (3). On each carbon atom, a pair of indices (atom ID, group ID) is shown for illustration purposes. The group IDs also correspond to Fig.4 in the main text. On the right hand side of each subfigure, the KS energy of the 1s electrons of carbon in the corresponding structure is plotted. The x-axis shows the KS energy with the Fermi level as a reference. The y-axis is a dummy variable for visualization. The atom IDs (group IDs) contribution of each peak is labeled by black font (white font). A zoom-in of the dense region in (d) is given for visualization.}\label{fig:CLS_1}
\end{figure*}

We calculated the Kohn-Sham (KS) energies $E$ of the 1s electrons of carbon in the ideal PAH structures (shown in Fig.4c-f in the main text) as a simple descriptor for core-electron spectroscopies. The calculation is done with density functional theory (DFT) using the FHI-aims package\cite{FHIaims}. We utilized the PBE functional\cite{PBE} in a tier 2 basis set with a 'tight' integration grid, please refer to the original package \cite{FHIaims} for the documentation about the parameters. Mulliken analysis is then performed to assign the KS energy $E$ of the 1s electrons of carbon in each PAH structure. The correlation of the grouping and these KS energies is demonstrated in Fig.~\ref{fig:CLS_1}.

As mentioned in the main text Sec.3A, groups 1-4 correspond to variations of carbon atoms with C coordination 2, while groups 5-7 with C coordination 3. Such significant difference is reflected in the gap between peaks of groups 5,6 and that of groups 1-4, as shown in Fig.~\ref{fig:CLS_1}.

This most prominent difference would already be recognized by a coarse classification using a large MSC bandwidth. With e.g. the smaller heuristic bandwidth described in the main text, further groups are distinguished, namely groups 1-4 and groups 5-7. As apparent, these groups indeed also exhibit different KS energies, i.e. there is a correlation between local atomic environment and physical property. This correlation is not linear though, as e.g. groups 5 and 6 in phenanthren (Fig.~\ref{fig:CLS_1}d) exhibit quite distinct KS energies, while groups 1 and 2 in naphtalene, anthracene and tetracene (Fig.~\ref{fig:CLS_1}a-c) have very small KS energy differences even though the distance between groups 1 and 2 in the two-dimensional MDS space is about the same as between groups 5 and 6 (cf. Fig. 4a in the main text). We also see from Fig.~\ref{fig:CLS_1}d, that the very subtle geometric differences within group 1 and within 2 are not properly resolved in phenanthren (Fig.~\ref{fig:CLS_1}d) in the heuristic two-dimensional MDS dimension. Higher embedding dimensions $S'$ together with an appropriate bandwidth $\delta_\text{MSC}$ would be necessary to further resolve these two groups into the four KS energy groups seen in Fig.~\ref{fig:CLS_1}d. As such and as discussed in the main text, the suitable resolution of the classification depends on the intended application.

\clearpage
\section{Performance with other SOAP settings}\label{sec:SOAP}
While performance of different SOAP settings is well studied\cite{cheng2020SOAP_heuristics} and can be systemically optimized\cite{SOAP_GAS}, we further demonstrate the framework performance with SOAP settings that differ from the heuristics mentioned the main text. In this section, we only change the SOAP settings applied on the Pd surface training set in Section 3B, the other heuristics are unchanged. Our default settings of double-SOAP are listed in the main text Sec.3. As a recap: the cut-off radii are $r^{\text{short}}_{\rm SOAP, cut}=3.320$\,{\AA} and $r^{\text{long}}_{\rm SOAP, cut} = 5.132$\,{\AA} for the Pd surfaces and the Gaussian width $\sigma^{\rm short/long}_{\rm SOAP} = {r^{\rm short/long}_{\rm SOAP, cut}} /8$. The maximum degrees of basis functions are $n^{\rm short}_\text{SOAP,max} = 8, l^{\rm short}_\text{SOAP,max} = 4, n^{\rm long}_\text{SOAP,max} = 4,$ and $l^{\rm long}_\text{SOAP,max}= 3$.

\subsection{Varying maximum degrees of basis functions in SOAP}\label{subsec:SOAP_1}
\begin{figure*}[h]
\centering
\includegraphics[width=0.8\textwidth]{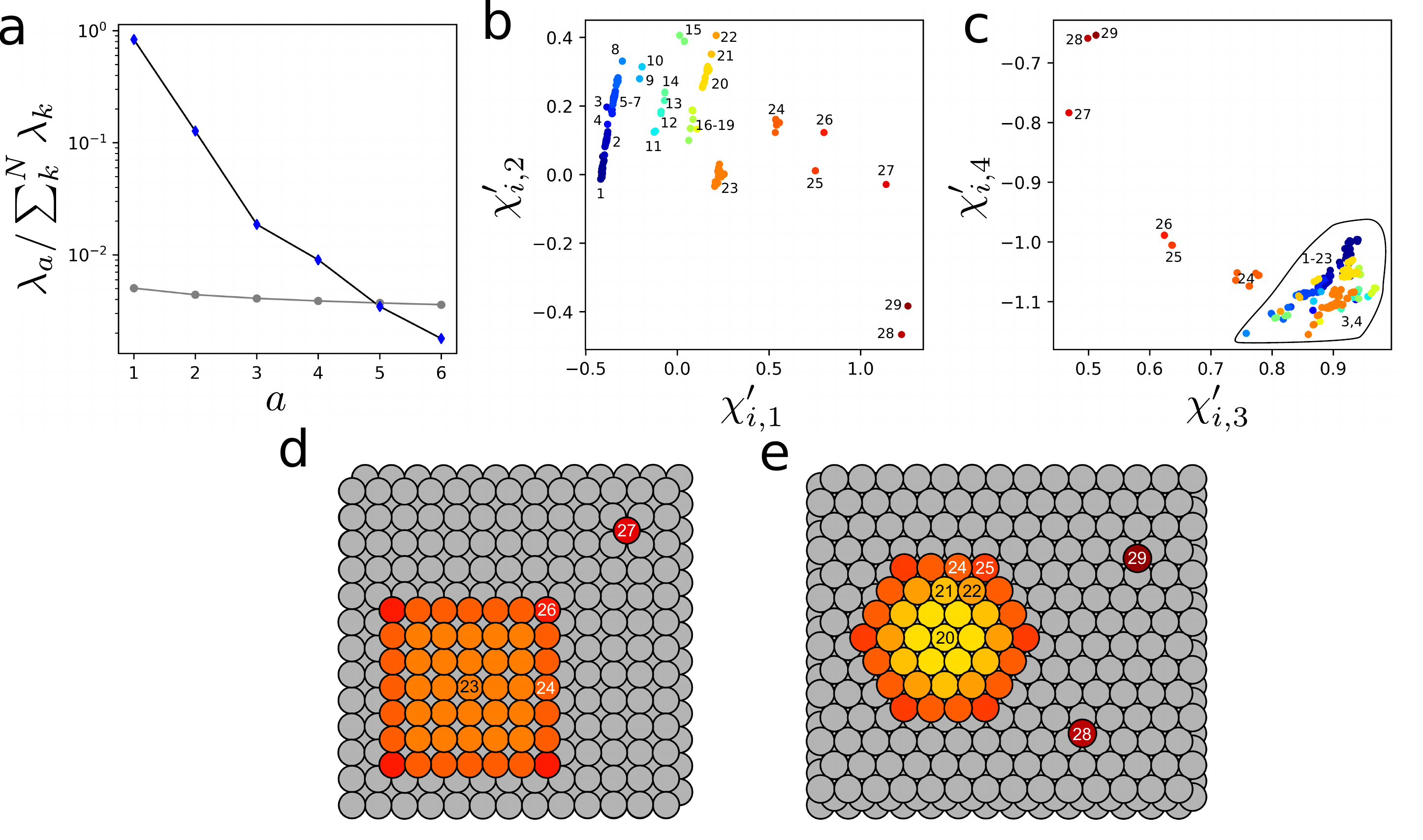}
\caption{The performance on basis of a SOAP descriptor with the maximum degrees of the basis functions increased by a factor of $\sim 1.5$. (a) The black line shows the normalized eigenvalues $\lambda_a/\sum_k^N \lambda_k$ of the Gram matrix in descending order, cf. Eq.2 for the crystalline Pd surface structure set (d-e). The grey line shows the broken-stick series, following Eq.5, $S'$ is estimated as 4. (b-c) Projection of the dataset with $N=1576$ points onto the planes spanned by the 1st/2nd and 3rd/4th MDS dimensions. The coloring of the points corresponds to an MSC clustering with a bandwidth of $\delta_{\rm MSC} = 0.0408$ as determined by the simple heuristics, see the main text. In a total of 29 equivalent atom classes are identified. (d-e) Top view of the atomic arrangement of the two nanostructured surfaces contained in the crystalline Pd surface structure set. In both cases, groups of atoms discussed are highlighted with color according to the MSC classification in (b). For clarity, we restrict this coloring to the island atoms and the adatoms, and the corresponding class index is shown only once in each structure.}\label{fig:SOAP_deg}
\end{figure*}

We first increase the maximum degrees of the basis functions to roughly 1.5 times of the default: $n^{\rm short}_\text{SOAP,max} = 12, l^{\rm short}_\text{SOAP,max} = 6, n^{\rm long}_\text{SOAP,max} = 6,$ and $l^{\rm long}_\text{SOAP,max}= 5$. The dimensionality of a double-SOAP vector increases from 220 to 672 due to the increase in the maximum degree of basis functions. The classification result is shown in Fig.~\ref{fig:SOAP_deg}.

The heuristics from the main text leads to an embedding dimension $S'=4$ and an MSC bandwidth $\delta_\text{MSC}=0.0408$. It results in a total of 29 groups of local environments. Please note that the group indices are assigned in order of the first MDS component $\mathbf{\chi}'_{i,1}$ and have no significance in the classification performance. Despite the huge increase in the SOAP dimensionality, the embedded SOAP $\mathbf{\chi}'$ resembles that of the default heuristics by a rotational transform. The feature of the first component representing the coordination number remains unchanged, which is not surprising as it is the quantity identified even by human intuition. Although the number of groups increases from 17 to 29 compared to the default SOAP settings, the grouping related to the islands as shown in Fig.~\ref{fig:SOAP_deg} is not strongly impacted (regarding the number of groups of corners, edges etc.). The group of inner atoms of an (111) island splits into three (group 20-22) and the adatoms on fcc/hcp sites are distinguished (group 28,29) with these SOAP settings. The other extra groups are of other subsurface atoms.

\subsection{Varying cut-off radii in SOAP}\label{subsec:SOAP_2}
\begin{figure*}[h]
\centering
\includegraphics[width=0.8\textwidth]{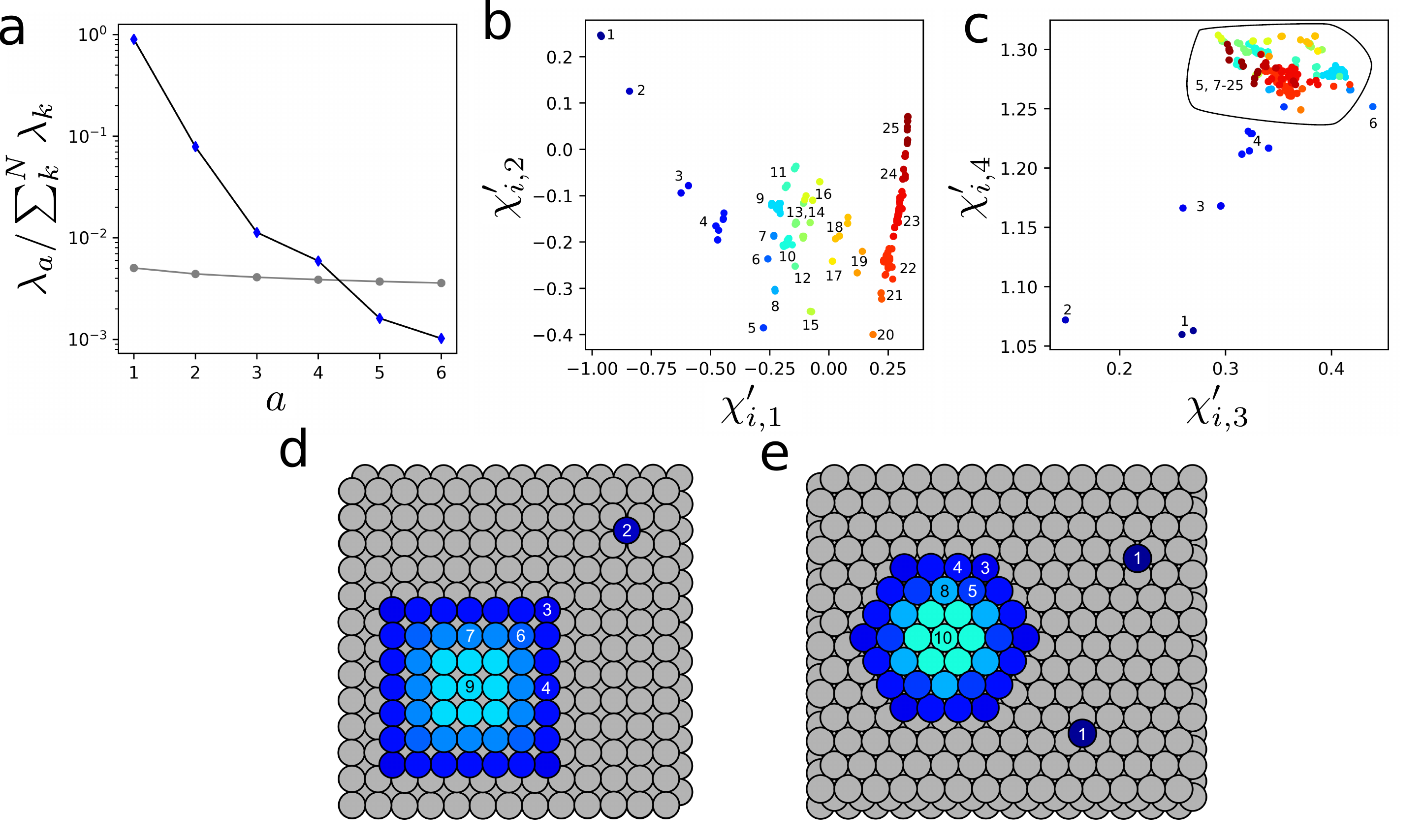}
\caption{The performance on the Pd training set with by a factor 1.5 increased cut-off radii for the SOAP descriptor compared to the default as described in the main text. (a) The black line shows the normalized eigenvalues $\lambda_a/\sum_k^N \lambda_k$ of the Gram matrix in descending order, cf. Eq.2 for the crystalline Pd surface structure set (d-e). The grey line shows the broken-stick series, following Eq.5, $S'$ is estimated as 4. (b-c) Projection of the data set with $N=1576$ points onto the planes spanned by the 1st/2nd and 3rd/4th MDS dimensions. The coloring of the points corresponds to an MSC clustering with a bandwidth of $\delta_{\rm MSC} = 0.0548$ as determined by the simple heuristics, see text. In a total of 25 equivalent atom classes are identified. (d-e) Top view of the atomic arrangement of the two nanostructured surfaces contained in the crystalline Pd surface structure set. In both cases, groups of atoms discussed are highlighted with color according to the MSC classification in (b). For clarity, we restrict this coloring to the island atoms and the adatoms, and the corresponding class index is shown only once in each structure.}\label{fig:SOAP_cut}
\end{figure*}

Besides varying the maximum degrees of basis functions, the cut-off radii $r^\text{long/short}_\text{SOAP,cut}$ may be of concern. Here we keep all other hyperparameters unchanged except increasing the cut-off to 1.5 times of the default settings. Specifically, $r^{\text{short}}_{\rm SOAP, cut}=4.979$\,{\AA} and $r^{\text{long}}_{\rm SOAP, cut} = 7.697$\,{\AA}, with the Gaussian width $\sigma^{\rm long/short}_{\rm SOAP} = {r^{\rm short/long}_{\rm SOAP, cut}} /8$. The result is shown in Fig.~\ref{fig:SOAP_cut}.

The heuristics from the main text leads to an embedding dimension $S'=4$ and a MSC bandwidth $\delta_\text{MSC}=0.0548$. It results in a total of 25 groups of local environments. Once again, the embedded SOAP $\mathbf{\chi}'$ largely resembles that of the default heuristics despite the huge increase in cut-off radii $r^{\text{long/short}}_{\rm SOAP, cut}$. The main difference in clustering is splitting the island atoms on (100) into three groups (6,7,9) and on (111) into another three (5,8,10).

Overall when both significantly increasing the maximum degrees of basis functions or cut-off radii, only a mild impact on the clustering results is obtained. This demonstrates that the default heuristics frequently employed in SOAP related works are fully adequate for the demonstrated Pd surface case. Of course, the parameters may be specifically optimized\cite{SOAP_GAS}, if this makes sense for a specific application or one may directly use excessive SOAP parameter settings. Unless the algorithm is to be executed at high frequency, the impact on the computation time is negligible.

\clearpage
\section{Bandwidth $\delta_\text{MSC}$ heuristics with different $\sigma_\text{smear}$}\label{sec:sigma}
\begin{figure*}[h]
\centering
\includegraphics[width=0.8\textwidth]{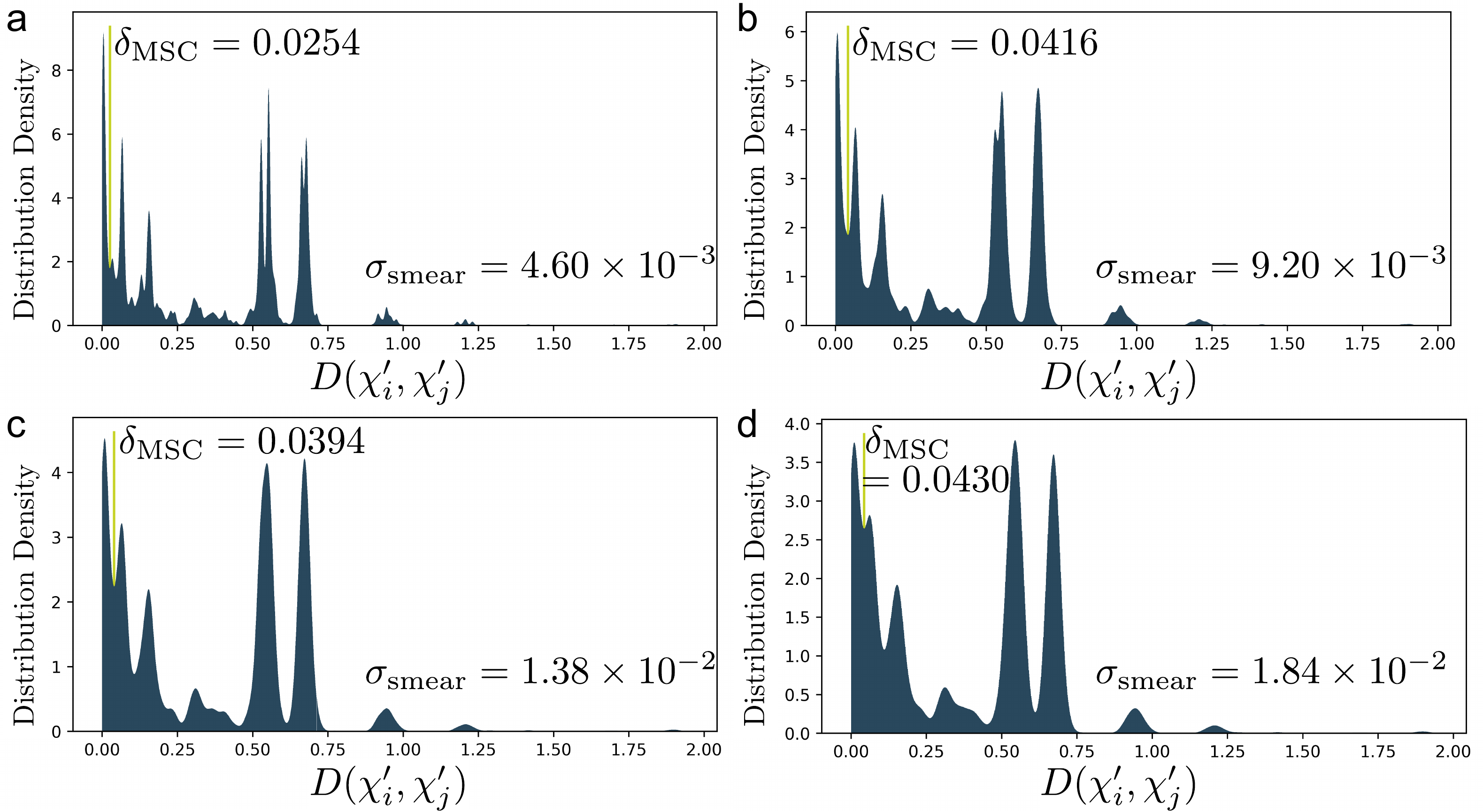}
\caption{The density of pairwise distances between embedded data points is shown. The yellow vertical lines show the locations of the first minimum, which is the $\delta_\text{MSC}$ chosen. The values of the smear parameter $\sigma_\text{smear}$ and bandwidth $\delta_\text{MSC}$ are listed in Tab.~\ref{tab:sigma}}\label{fig:sigma}
\end{figure*}

As discussed in the main text, the heuristics of estimating the MSC bandwidth $\delta_\text{MSC}$ comes from locating the first minimum in the distribution of pairwise distances $D(\mathbf{\chi}'_i,\mathbf{\chi}'_j)$ between the embedded SOAP vectors. This distribution is approximated using a Gaussian Kernel Density Estimator with a width $\sigma_\text{smear}$. The classification applied on the Pd training set, with values besides the default value ($9.20\times10^{-3}$) are tabulated in Tab.~\ref{tab:sigma}. The other settings of SOAP and embedding dimension $S'$ remain the same as in Sec.3B. The distribution densities are shown in Fig.~\ref{fig:sigma}.

\begin{table}[h]
\begin{center}
\caption{Number of identified equivalence classes for the Pd surface structure set, with different smearing parameters $\sigma_\text{smear}$, while maintaining the heuristic determination of the MSC bandwidth $\delta_{\rm MSC}$ described in Section 2C. The number of 17 classes resolved for $\sigma_\text{smear} = 9.20\times10^{-3}$ was the case discussed in Section 3B.}\label{tab:sigma}
\begin{ruledtabular}
\begin{tabular}{ccc}
$\sigma_\text{smear}/10^{-3}$ & $\delta_{\rm MSC}$ & No. of identified classes \\
\hline
4.60  & 0.0254 &25 \\
9.20  & 0.0416 &17\\
13.8  & 0.0394 &17\\
18.4  & 0.0430 &16\\
\end{tabular}
\end{ruledtabular}
\end{center}
\end{table}

As expected, a larger $\sigma_\text{smear}$ results in a more coarse-grained distribution. An extremely small $\sigma_\text{smear}$ (e.g. half of our default $9.20\times10^{-3}$) leads to a significantly smaller bandwidth $\delta_\text{MSC}$ and more classes. On the other hand, beyond the default value (up to doubling the default, $1.84\times10^{-2}$), the number of groups is steady in this test case. From this demonstration, we are confident that the given heuristics reasonably estimate a suitable starting point of bandwidth $\delta_\text{MSC}$. Furthermore, users may utilize this smoothened distribution to guide the fine tuning of $\delta_\text{MSC}$ according to other typical length scales represented as minima in the distributions.

\clearpage
\section{Out-of-sample Classification on Data from Molecular Dynamics Simulation}\label{subsec:MD}
\begin{figure*}[h]
\centering
\includegraphics[width=0.4\textwidth]{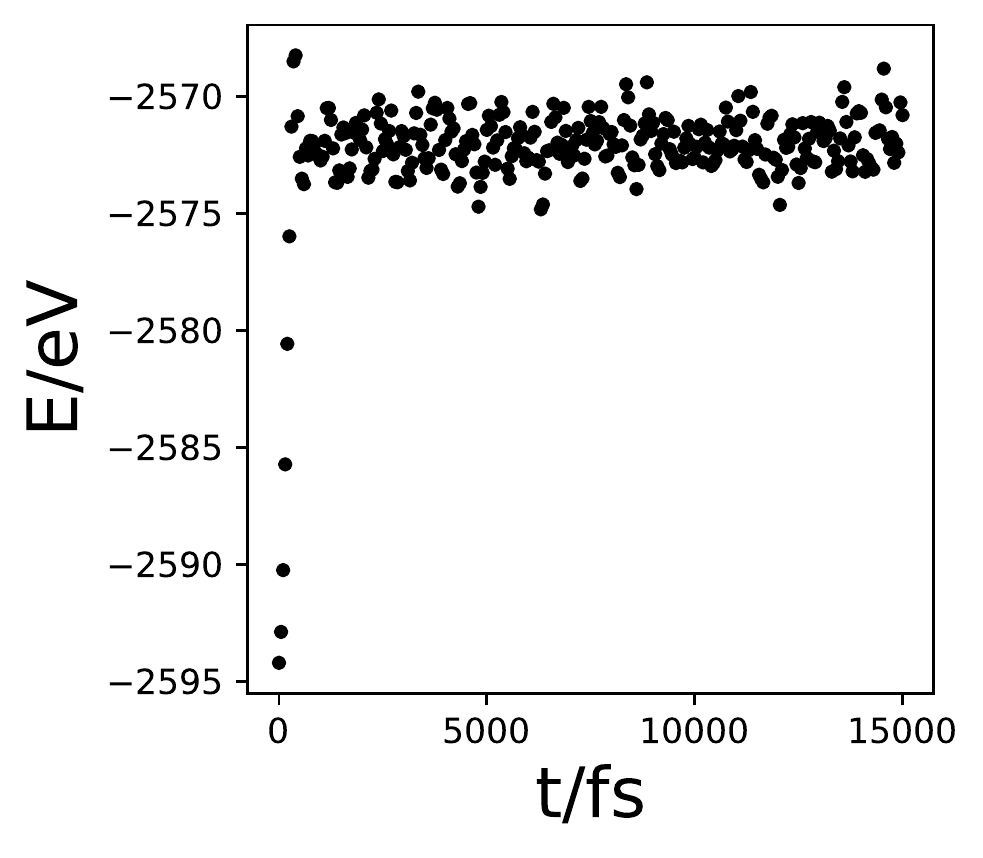}
\caption{Time evolution of total energy $E$ output from the MD simulation of the structured Pd(100) surface shown in Fig.5a in the main text. Equilibrium is clearly reached by $2\times 10^3$fs. Please note that data points are only plotted every 100 time steps.}\label{fig:MD_Eqm}
\end{figure*}

In this section, we further explore the out-of-sample classification function. We first prepare a test set as out-of-sample data using molecular dynamics (MD) simulation applied on the structured Pd (100) surface as shown in Fig.5a in the main text. Specifically, the simulations have been conducted using LAMMPS\cite{lammps}, with a timestep of 0.5\,fs and an embedded atom potential\cite{PdEAM}. A Nose-Hoover thermostat was employed for sampling an $NVT$ ensemble at 300K. Equilibrium has been reached by 2000\,fs as shown in Fig.~\ref{fig:MD_Eqm}.

\begin{figure*}[h]
\centering
\includegraphics[width=0.8\textwidth]{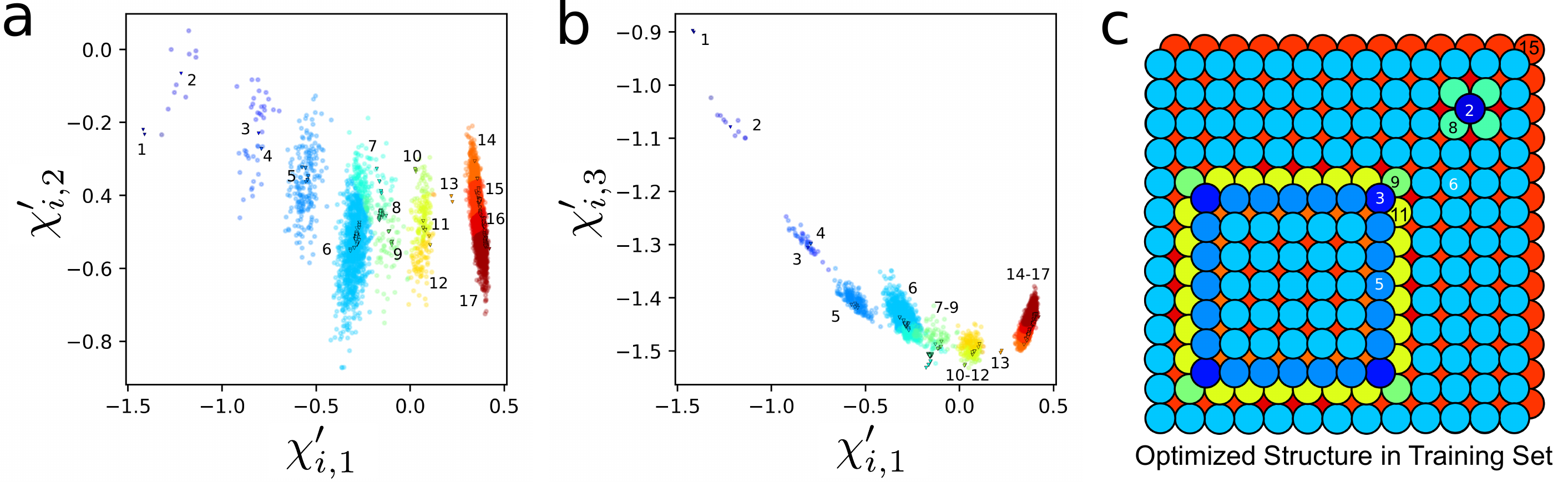}
\caption{(a-b) The three dimensional MDS space obtained in Sec.3B, with the dots representing MD (out-of-sample) data. Upside down triangles represent data points from the training data set. The color corresponds to the group index. (c) A reference figure of the optimized structure in the training set, in full-colors according to the group indices.}\label{fig:MD_MDSSpace}
\end{figure*}

Between $t=3\times 10^3$fs to $1.5\times 10^4$fs, 10 snapshots are randomly selected. The SOAP vector of each atom in these 10 snapshots is computed with exactly the same settings in Sec.3B. All these SOAP vectors from the MD $\mathbb{\chi}_\text{new}$ are taken as out-of-sample data and embedded into the MDS space obtained from the Pd dataset in section 3B. When performing out-of-sample classification, we assign each $\mathbb{\chi}'_\text{new}$ to its nearest cluster center. This is an alternate strategy besides the one demonstrated with the nanoparticle in the main text. The result of the clustering is shown in Fig.~\ref{fig:MD_MDSSpace}.

The fraction of atoms which remain in their initial group (in the training data set) is tabulated in  Tab.~\ref{tab:MD}. A fraction of $90\%$ or more of the adatoms, island edge or simple surface atoms remain in the same environment class at 0K and 300K, despite the thermal displacements. On the other hand, the corner atoms have a $50\%$ probability of leaving their 0K class due to their larger anisotropic vibrations. For the future development, an improvement can be made with exploiting the correlations in the classification of the individual atoms in successive snapshots along the trajectory.
\clearpage
\begin{table}[h]
\begin{center}
\caption{The fraction of atoms in MD snapshots with group index same as that in the initial (training set) structure. The group indices not listed here are those absent groups in the optimized Pd(100) surface.}\label{tab:MD}
\begin{ruledtabular}
\begin{tabular}{cccc}
Group index& Matching$\%$ & Expected coordination no. & Description \\
\hline
2 & 90.0$\%$ & 4 & Adatom \\
3 & 50.0$\%$ & 6 & Corner atom \\
5 & 99.5$\%$ & 7 & Edge atom \\
6 & 89.0$\%$ & 8 & Surface atom \\
8 & 40.0$\%$ & 9 & Surface atom below an adatom \\
9 & 90.0$\%$ & 9 & Surface atom below a corner \\
11 & 51.7$\%$ & 10 & Surface atom below an edge \\
14 & 50.0$\%$ & 12 & Bulk Group \\
15 & 50.9$\%$ & 12 & Bulk Group \\
16 & 35.5$\%$ & 12 & Bulk Group \\
17 & 78.2$\%$ & 12 & Bulk Group \\
\end{tabular}
\end{ruledtabular}
\end{center}
\end{table}

\bibliography{aipsamp_si}